\documentclass[12pt]{article}
\usepackage{amsfonts}
\usepackage{bm}
\usepackage{amsmath}
\usepackage{amscd}

\usepackage{amssymb,lscape}

\allowdisplaybreaks

\numberwithin{equation}{section}

\newcommand{\bmat}{\left(\begin{array}}
\newcommand{\emat}{\end{array}\right)}

\def\gtrsim{\mathrel{\raise.3ex\hbox{$>$\kern-.75em\lower1ex\hbox{$\sim$}}
}
}

\def\-{\hphantom{-}}

\def\s2{\frac{1}{\sqrt2}}

\def\beq{\begin{equation}}
\def\eeq{\end{equation}}
\def\beqa{\begin{eqnarray}}
\def\eeqa{\end{eqnarray}}

\def\tM{{\tilde M}}
\def\tL{{\tilde L}}

\def\mg{m_{3/2}}
\def\mg2{m^2_{3/2}}

\def\Dsl{\,\raise.15ex\hbox{/}\mkern-13.5mu D} 

\def\be{\begin{equation}}
\def\ee{\end{equation}}
\def\bea{\begin{eqnarray}}
\def\eea{\end{eqnarray}}

\DeclareMathOperator{\GL}{\mathit{GL}}

\DeclareMathOperator{\Es7}{\mathit{E}_{7(7)}}

\newcommand{\nn}{\nonumber}

\begin{document}
\pagestyle{plain}
\begin{titlepage}
\rightline{\small IPhT-T13/227}
\begin{center}
\large{\bf The gauge structure of Exceptional Field Theories \\ and the tensor
hierarchy\\[4mm]}
\large{  G. Aldazabal${}^{a,b}$, M. Gra\~na${}^c$, D.
Marqu\'es${}^d$ , J. A. Rosabal${}^{a,b}$
 \\[4mm]}
\small{
${}^a${\em Centro At\'omico Bariloche,} ${}^b${\em Instituto Balseiro
(CNEA-UNC) and CONICET.} \\[-0.3em]
{\em 8400 S.C. de Bariloche, Argentina.}\\
[0.3cm]
${}^c${\em Institut de Physique Th\'eorique,
CEA/ Saclay \\
91191 Gif-sur-Yvette Cedex, France}  \\
[0.3cm]
${}^d${\em Instituto de Astronom\'ia y F\'isica del Espacio
(CONICET-UBA)} \\
C.C. 67 - Suc. 28, 1428 Buenos Aires, Argentina.  \\[0.2cm]
{\verb"aldazaba@cab.cnea.gov.ar , mariana.grana@cea.fr"}\\
{\verb"diegomarques@iafe.uba.ar , rosabalj@ib.cnea.gov.ar"}\\
[0.5cm]}
\small{\bf Abstract} \\[0.3cm]
\end{center}

We address the construction of manifest U-duality invariant generalized
diffeomorphisms. The closure of the algebra requires an extension of the
tangent space to include  a
tensor hierarchy  indicating the existence of an underlying unifying structure,
compatible with $E_{11}$ and Borcherds algebras constructions.
We begin with four-dimensional gauged maximal supergravity, and build a
generalized Lie derivative  that encodes all the gauge transformations of the
theory. A generalized frame is introduced, which accommodates for all the
degrees of freedom, including the tensor hierarchy. The generalized Lie
derivative defines generalized field-dependent fluxes containing all the
covariant quantities in the theory, and the closure conditions give rise to
their corresponding Bianchi Identities.
We then move towards the construction of a full generalized Lie derivative
defined on an extended space, analyze the closure conditions, and explore the
connection with that of maximal gauged supergravity via a generalized
Scherk-Schwarz reduction, and with $11$-dimensional supergravity.

\vfill


\end{titlepage}


{\footnotesize \tableofcontents}

\section{Introduction}\label{section:Intro}

Incorporating stringy symmetries like T-duality into a field theory, or into a
(generalized) geometric description led to Double Field Theory (DFT)
\cite{Siegel:1993th,DFT} (see \cite{reviewDFT} for reviews and further
references) or Generalized Geometry (GG) \cite{GG,WaldramR} (see
\cite{reviewGG} for a review).
T-duality invariance is realized in DFT by doubling the
coordinates of the
internal  $n$-dimensional compactification space.
Namely,  besides the usual coordinates  conjugate to compact momentum in string
toroidal compactifications, a new set
of coordinates  conjugate to string windings is
included.
The so-called section condition (or strong
constraint) restricts the fields to depend only on half of the double coordinates.
In GG the coordinates themselves are not doubled,
but one considers
generalized vectors living on
a generalized tangent space with twice the dimension of the ordinary tangent
space.
Thus, for $n$ compact dimensions, vector fields in DFT or GG span a vector
representation of the full T-duality group
$O(n,n)$.  A positive definite metric on this space  can be defined in terms of
 the massless states in the NSNS sector of the superstring. Enlarging the
diffeomorphism symmetry to include the gauge transformation of the two-form
lead to consider a generalized diffeomorphic transformation, encoded in a
generalized Lie derivative \cite{Siegel:1993th,TGG,DFT}.

More generally, promoting U-duality to a symmetry requires a further extension
of
the double tangent space into an extended or exceptional generalized space
\cite{HullM,PachecoWaldram}, or in
the spirit of DFT
enlarging the compact space itself into  a mega-space  (a mega-torus
\cite{Dall'Agata:2007sr,Hillmann:2009ci,Aldazabal:2010ef} in
the case of toroidal backgrounds) with  derivatives  spanning a
representation of the U-duality group.
The U-duality symmetry groups in question are the exceptional groups
$E_{n+1(n+1)}$ of
toroidal compactifications, where $n$ is the dimension of the  compactification
space in String Theory ($n+1$ in M-theory). An internal positive definite
metric  can be defined and
parameterized in terms of the degrees of freedom of Type II or M-theory. In
this case the diffeomorphisms and gauge transformations are encoded in an
extended generalized Lie derivative \cite{WaldramE,Berman:2012vc}.

Interestingly enough, in DFT or GG it is possible to include  not only the
symmetries
corresponding to the $n$ compact dimensions, but also those of the $d$
space-time dimensions.
Namely, the full tangent space is doubled, and full $O(D,D)$ ($D=n+d$)
covariant generalized diffeomorphisms can be constructed. This proves to be a useful unified
description, where afterwards, the $d$-dimensional space-time can be
decompactified, amounting in DFT to disregard the dual space-time coordinates,
 leaving a theory with
$GL(d)\times O(n,n)$ symmetry. In
particular, after a generalized Scherk-Schwarz reduction, such theories were
shown to lead to the electric bosonic sector of $d$-dimensional half-maximal
gauged
supergravities \cite{Aldazabal:2011nj}.

U-duality $E_{n + 1(n+1)}$ invariant constructions at the full $D$-dimensional
level are tricky. The simplest setups consider only the internal sector, and
therefore correspond to truncations of a full Exceptional Generalized Geometry
or Exceptional Field Theory (EFT). Restricted to this sector, constructions of
generalized Lie derivatives for dimensions $n\leq 6$ can be found in
\cite{WaldramE,Berman:2012vc}, and invariant actions in \cite{Berman}. The $n = 7$  case is discussed in
\cite{Godazgar:2013rja,SC}, and requires the introduction of the
$11$-dimensional
dual graviton. For dimensions $n \geq 8$, the groups in question are not even
finite. A  description of gauged maximal supergravities in $d=2$ with $n=8$
covariance is available in \cite{SamtlE9}. The situation is far less clear for
$n>8$, though there are indications pointing to an $E_{11}$ underlying
structure \cite{E11,west} or Borcherds algebras-based constructions
\cite{Palmkvist:2011vz}-\cite{Henneaux:2010ys}. Some of these
constructions for $n\leq 6$ were explored in the context of generalized
Scherk-Schwarz compactifications in \cite{Berman:2012uy,Musaev:2013rq,nosotros},
 related to generalized geometric notions in
\cite{WaldramE,nosotros,Park:2013gaj,Cederwall:2013naa}, and related to
F-theory \cite{Blair:2013gqa}.

A key question is how to couple the internal sector discussed above with the
external $d$-dimensional space-time. Previous works in this direction are
\cite{KKandWinding,Ugravity} and more recently an $E_{6(6)}$ invariant
EFT was presented in  \cite{HohmSamt}. In the present
article we perform a step forward towards a unified description of gauge
transformations in terms of a generalized Lie derivative.

We begin with the
simpler setup of generalized Scherk-Schwarz reductions in $d = 4$ and $n = 6$
which lead to $4$-dimensional gauged ${\cal N} = 8$ supergravity \cite{de
Wit:2007mt}, in which the
gaugings \cite{deWit:2002vt} are obtained from twists of an internal
$56$-dimensional
parallelizable mega-space \cite{nosotros}. The methodology can be easily
adapted to
other groups and dimensions.
The extended Lie derivative that encodes all the gauge transformations of the
reduced theory  is one of the central results  of
the present article. It is constructed from a careful study of closure of
the extended diffeomorphism algebra, and it requires extending the $4+56$
tangent space into a larger $E$-tangent space that accommodates all the
$p$-form hierarchy \cite{Weidner}-\cite{Bergshoeff:2009ph}. We pay special attention to the role played in the closure of the diffeomorphism algebra by the so-called intertwining tensors, built out of the embedding tensor.

A generalized vector on the full $E$-tangent space contains $4$ components that
generate diffeomorphisms in the ``external'' space, $56$ components that
generate diffeomorphisms in the generalized ``internal'' space (which contain
the gauge transformations of the vector fields), and extra components that
generate the gauge transformations of the $p$-form fields in maximal gauged
supergravity.  A
field-dependent generalized frame for a generalized $E$-vector can then be
parameterized by a $4$-dimensional bein, $56$ one-form gauge fields, the scalar
coset matrix, and extra space-time $p$-forms that carry internal indices in the
modules of the so-called tensor hierarchy. By evaluating the generalized Lie
derivative on generalized frames, a set of
extended
dynamical  fluxes can be derived. We show that these fluxes contain all form
field strengths of $4$-dimensional gauged  maximal supergravity, and agree
with those found in the breaking of $E_{11}$ into $\GL(4) \times \Es7$
\cite{west}. Moreover, the closure
conditions for the generalized fluxes reproduce the Bianchi Identities for the
curvatures.

Based on the lessons learned in the compactified case, we then move to the
general case with no compactification ansatz assumed. Here, we begin general
and do not specify the U-duality group. For any U-duality group $E_{n+1(n+1)}$, the generalized coordinates contain $d$ ``external'' components $x^\mu$, and the $n+1$ ``internal'' coordinates are embedded in a given representation of $E_{n+1(n+1)}$, to achieve duality covariance. The distinction
between internal and external is only formal since no compactification is assumed. At any stage a section condition can be imposed that selects an $n+1$ section of the generalized internal space, allowing to make contact with the $d + n + 1=11$-dimensional space-time of
$11$-dimensional supergravity, or $d+n=10$ of type II theories. We see that the intertwining
(embedding) tensors uplift to differential intertwining operators and analyze
the role played by them in closure. We comment on the
relation between the full
generalized Lie derivative, and the connection to the $4$-dimensional one upon
a generalized Scherk-Schwarz reduction. Intriguingly, we identify seemingly
obstructions in the construction
of fully covariant generalized diffeomorphisms, and suggest how to circumvent
them for the different U-duality groups.

The paper is organized as follows. In Section \ref{sec:summary} we present the
setup and the main results of the paper. In Section \ref{sec:E} we show how to
include the tensor hierarchy in the generalized Lie derivative for the reduced
theory,  and analyze the role of the intertwiners and $p$-forms in the closure
of the gauge algebra. We also give the explicit form of the generalized frame,
parameterized in terms of the degrees of freedom of gauged maximal
supergravity, and show that we reproduce the corresponding gauge transformations
by acting with the generalized diffeomorphism. This section includes the
computation of generalized fluxes and Bianchi Identities.
In Section \ref{sec:extendedst}, we give a first step towards the construction
of a universal (namely, valid for any exceptional duality group) covariant  full
generalized Lie derivative in the extended space, and explore the closure of
the algebra. Concentrating on the case of $E_{7(7)}$, we make contact in
Section \ref{sec:11dsugra} with 11-dimensional supergravity by breaking
$E_{7(7)}$ into $SL(8)$ and then further $GL(7)$, where the latter acts on the
ordinary ``internal" tangent space. We first show how the fields assemble into
$E_{7(7)}$ representations. As we go up in the tensor hierarchy, we need to go
further and further beyond supergravity and include more and more non-geometric
objects (``U" or ``exotic" branes \cite{Hullexotic}-\cite{exoticbranes}) to fill
up $E_{7(7)}$ representations. We then restrict to conventional 11-dimensional
supergravity, construct the generalized frame and recover from its
generalized Lie derivative the gauge transformations of 11-dimensional supergravity. We
conclude in Section
\ref{sec:Conclu}.

\section{Setup and summary of main results} \label{sec:summary}

In this section  we briefly review some of the main results of this paper.
First we
present some developments that appeared recently in the
literature, related to generalized diffeomorphisms in the internal space, which
serve for a base to the extensions considered in this article. Then we
summarize our  main results.
\subsection{Summary of previous results}

Recently, U-duality covariant generalized diffeomorphisms for the internal
sector of maximal supergravity were considered in \cite{WaldramE} and
\cite{Berman:2012vc}. In order to realize manifest $E_{n+1(n+1)}$ invariance,
the internal space can be extended to coincide with the dimension of a given
representation of the U-duality group.

 Noting the internal derivatives as $\partial_M$,  the covariant exceptional or
extended
Dorfman bracket takes the general structure
\be
( {\cal L}_{\hat \xi} \hat V)^M  = \hat \xi^P \partial_P \hat V^M - \hat  V^P
\partial_P \hat \xi^M + Y^M{}_N{}^P{}_Q \partial_P \hat \xi^Q \hat V^N
\label{extendedDorfman}
\ee
where $\hat \xi$ and $\hat V$ are vectors in the same representation than the
coordinates. We are using the notation that hatted quantities depend on all the
internal and external coordinates, although at this stage the external
dependence is not important. This generalized Lie
derivative  is the internal one, and has to be distinguished from the full
generalized Lie derivative to be constructed later, which will be hatted. The
tensor $Y$ depends on the particular U-duality group invariants, and measures
the
deviation from standard Riemannian geometry governed by the first two terms
above, which correspond to the Lie derivative in the internal sector. We give
its components for $E_{n+1(n+1)}$ with $n\le5$ in (\ref{Ytensors}), and  for
$E_{7(7)}$ in (\ref{Ytensor}).

Closure of the generalized Lie derivative is not automatic,
and requires imposing closure constraints. These constraints can be solved
by imposing a ``section condition" that restricts the
theory further \cite{WaldramE}\footnote{ The section condition states that $P^{MN}{}_{PQ}
\partial_M \otimes \partial_N =0$ (where $P$ is the projector to the second module of the duality group) must annihilate any field or gauge parameter, and also products of them.}. This condition implies that the
following two operators vanish when acting on any product of fields and/or
gauge parameters
  \bea
  Y^M{}_P{}^N{}_Q\ \partial_M \otimes \partial_N  &=& 0 \label{sec1} \\
  (Y^M{}_Q{}^N{}_P Y^P{}_R{}^T{}_S - Y^M{}_R{}^N{}_S \delta^T_Q)\ \partial_{(N}
\otimes \partial_{T)}  &=& 0\label{SectionCondition}
  \eea
In the paper we will not assume these constraints, but they can be implemented at any stage.

Closure requires actually more relaxed constraints. In particular, in the
context of generalized Scherk-Schwarz configurations
\cite{Berman:2012uy},\cite{Musaev:2013rq},\cite{nosotros}, the section condition was proved to be too strong \cite{Grana:2012rr}, and only a subset
of gauged supergravities can be obtained upon dimensional reduction when the
framework is restricted by it. In contrast, closure constraints allow for
solutions that violate the strong constraint,  which permit to make contact
with all the
admissible deformations of the theory (see for example
\cite{Dibitetto:2012rk,Berman:2012uy,Musaev:2013rq,nosotros} for more details). Twisting the generalized Lie derivative (\ref{extendedDorfman})
with a U-duality valued twist matrix $U_A{}^M$ leads to the gaugings
$F_{AB}{}^C$
\be
F_{AB}{}^C = 2 \Omega_{[AB]}{}^C + Y^C{}_B{}^D{}_E \Omega_{DA}{}^E \ , \ \ \ \
\Omega_{AB}{}^C = U_A{}^M \partial_M U_B{}^N (U^{-1})_N{}^C\label{gaugings}
\ee
which automatically satisfy the linear constraints of gauged maximal
supergravity, projecting out the representations not allowed by supersymmetry.
In addition, the closure constraints force the gaugings to satisfy the quadratic
constraints of gauged maximal supergravity
\be
F_{AD}{}^E  F_{BE}{}^F - F_{BD}{}^E  F_{AE}{}^F + F_{AB}{}^E F_{ED}{}^F = 0
\label{quadraticconstraints}
\ee
While the antisymmetric part takes the form of a Jacobi Identity, the symmetric
part is not automatically satisfied and depends on the symmetric part of the
gaugings $F_{(AB)}{}^E$, called the intertwining tensor, which vanishes under
the following contraction
\be
F_{(AB)}{}^E F_{ED}{}^F = 0\label{symF}
\ee
due to the symmetrization of (\ref{quadraticconstraints}).

Let us now specialize to the $E_{7(7)}$ case, which will be the case we
explore in Section \ref{sec:E}.  In a previous paper \cite{nosotros} we have
addressed the construction of an extended geometry for the internal sector of
$4$-dimensional maximal gauged supergravity. In order to realize manifest
$E_{7(7)}$ invariance, the internal tangent space was taken to be
$56$-dimensional,
in accordance with the dimensionality
of the fundamental $\bf 56$ representation of $E_{7(7)}$. In this case, the
group must be augmented with an $\mathbb{R}^+$  factor,  necessary for closure
of the algebra, as explained in \cite{WaldramE}. There are two $E_{7(7)}$
invariants, a symplectic metric $\omega_{MN}$ (which raises and lowers indices)
and the projector to the adjoint $\bf 133$ representation $P{}^M{}_N{}^P{}_Q$.
In terms of them, the $Y$-tensor takes the form
\be
Y^M{}_N{}^P{}_Q = - 12 P^{MP}{}_{NQ} +  \frac 1 2
\omega^{MP}\omega_{NQ}\label{Ytensor}
\ee
Performing a twist in terms of an internal index-valued frame $\hat E_{\bar A}{}^M$,
we found the expression for the internal generalized fluxes generated by the
corresponding
mega-twisted-torus
\be
\mathbb{F}_{\bar A \bar B}{}^{\bar C} = 2 \Omega_{[\bar A \bar B]}{}^{\bar C} + Y^{\bar
C}{}_{\bar B}{}^{\bar D}{}_{\bar E}\ \Omega_{\bar D \bar A}{}^{\bar E} \ , \ \
\ \ \
\Omega_{\bar A \bar B}{}^{\bar C} = \hat E_{\bar A}{}^M \partial_M \hat E_{\bar
B}{}^N (\hat E^{-1})_N{}^{\bar C}
\ee
Although the above analysis was restricted purely to the internal sector, in
the particular case of a generalized Scherk-Schwarz compactification the frame
decomposes as
\be \label{SSansatz}
\hat E_{\bar A}{}^M(x,Y) = \Phi_{\bar A}{}^B(x) U_B{}^M(Y)
 \ee
with $\Phi_{\bar A}{}^B(x)$ containing the scalar fields and $U_B{}^M(Y)$ the
twist matrix that generates the gaugings in the reduced theory. Since the
scalars depend on the $4$-dimensional space-time coordinates $x^\mu$, with $\mu
= 1,\dots,4$, they were regarded as constants form the internal sector point of
view. Then, after a Scherk-Schwarz reduction, the above fluxes can be cast in
the form
\be
\mathbb{F}_{\bar A \bar B}{}^{\bar C}(x) = \Phi_{\bar A}{}^A \Phi_{\bar B}{}^B
(\Phi^{-1})_C{}^{\bar C} \ F_{AB}{}^C \label{gaugingsF}
\ee
where $F_{AB}{}^C$ are taken to be constant and identified with the gaugings of
maximal supergravity. It can be checked that by construction they belong to the
$\bf 56$ + $\bf 912$ representations, and then automatically satisfy the linear
constraints of the theory. Finally we note that we are distinguishing between
three types of indices: $M,N,\dots$ refer to curved internal indices in the
extended theory, $A,B,\dots$ refer to curved internal indices in the reduced
theory, and $\bar A, \bar B,\dots$ are the flat indices in both.
\subsection{Summary of new results}

In this paper we explore how to extend the above construction by coupling the
missing
 space-time dimensions. We will start with what we will call the
``compactified" case, in which we assume a Scherk-Schwarz-type ansatz, with
$Y$-independent fluxes of the form (\ref{gaugingsF}). This case leads to $4$-dimensional
maximal
gauged supergravity. Later, we will extend most of our results to the
``decompactified" case, i.e. where we assume that the generalized tangent space
splits into 4 and 56 (or actually more, as we will see) directions, but where
everything depends in a generic way on external and internal coordinates.

We begin with the compactified case, specializing  to the $E_{7(7)}$ U-duality
group, and explore the role of the intertwining tensors
and the tensor hierarchy in the closure of the algebra. We will extract lessons
that will help in  building a full generalized Lie derivative in
the extended space.

\subsubsection{Generalized diffeomorphisms in gauged maximal supergravity}

Let us begin with the $4$-dimensional case of gauged maximal
supergravity, with U-duality group $E_{7(7)}$. All the expressions found here
coincide with the results obtained
in the gauged maximal supergravity formulation of \cite{de Wit:2007mt}, the
$E_{11}$ approach in \cite{E11}-\cite{west} or Borcherds algebras constructions
\cite{Palmkvist:2011vz}-\cite{Henneaux:2010ys}. The generalized $E$-vector fields (in particular gauge
parameters) are of the form
\bea
\xi^{\mathbb{A}} &=& (\xi^\mu, \xi^{A}, \xi_\mu{}^{<AB>},
\xi_{\mu\nu}{}^{<ABC>},
\xi_{\mu\nu\rho}{}^{<ABCD>},\dots) \\
&=& (\xi^\mu, \xi^A, \xi_{\mu}{}^\alpha,\xi_{\mu\nu}{}^{\cal A},
\xi_{\mu\nu\rho}{}^{\bf A},\dots)\nn
\eea
where $A, B, C$ are indices in the first module of the duality
group (for $E_{6(6)}$ and $E_{7(7)}$, this corresponds to the fundamental representation), and $<\dots>$ means a projection from the tensor product of various
fundamental indices to a particular irreducible representation (or sums of
irreducible representations), labeled by $\alpha,{\cal A}, {\bf A},\dots$ on
the second line. In the case of $E_{7(7)}$,  $A$ is a fundamental $\bf 56$
index, $\alpha$ takes values
in the adjoint $\bf 133$ representation, $\cal A$ belongs to the $\bf 912$
representation, $\bf A$ to the $\bf 8645 + 133$ representation, and so on. We
will comment on the end of this hierarchy in due time. If we think of these as
gauge parameters, they include
the usual
Riemannian geometry diffeomorphisms generated by four-dimensional vectors
$\xi^{\mu}$,   (extended generalized) diffeomorphisms of the
internal space $\xi^{M}$, plus new extra gauge parameters required for the
gauge algebra to close. Note that these gauge parameters do not carry a hat,
because they only depend on the external coordinates. The general structure of
the generalized Lie derivative is given by
\be
\left(\hat {\cal L}_{\xi_1} \xi_2\right)^{\mathbb{A}} = \xi_1^{\mathbb{B}}
\partial_{\mathbb{B}} \xi_2^{\mathbb{A}} - \xi_2^{\mathbb{B}}
\partial_{\mathbb{B}} \xi_1^{\mathbb{A}}  +
W^{\mathbb{A}}{}_{\mathbb{B}}{}^{\mathbb{C}}{}_{\mathbb{D}}\partial_{\mathbb{C}}
 \xi_1^{\mathbb{D}} \xi_2^{\mathbb{B}} + F_{\mathbb{BC}}{}^{\mathbb{A}}
\xi_1^{\mathbb{B}} \xi_2^{\mathbb{C}}\label{reducedgeneralform}
\ee
where, since we are considering the compactified case here, we have
$\partial_{\mathbb{A}} = (\partial_\mu, 0 ,\dots)$. Here, we put a hat on the
generalized Lie derivative to emphasize that it corresponds to the
(compactified) full Lie derivative. In the core of the paper we will give more
explicit expression for all these quantities, here we are simply stating the
general form of our results. The
$W^{\mathbb{A}}{}_{\mathbb{B}}{}^{\mathbb{C}}{}_{\mathbb{D}}$
tensor is formed by $GL(4)$ and $E_{7(7)}$ invariants, and its purely external
components vanish in accordance with Riemannian geometry. While the first three
terms are un-gauged, the tensor $F_{\mathbb{BC}}{}^{\mathbb{A}}$ depends
linearly on the gaugings, and then carries the information on the internal
$Y$-tensor introduced in (\ref{extendedDorfman}) through (\ref{gaugings}).

Out of the gaugings and generators of the group, one builds a hierarchy of
intertwining tensors\footnote{ For example, $F_{\alpha}{}^A =
(t_{\alpha}){}^{BC} F_{(BC)}{}^A$.}
\be
F_{\alpha}{}^A\ , \ \ \ F_{\cal A}{}^{\alpha} \ , \ \ \ F_{\bf A}{}^{\cal A} \
, \ \ \ \dots
\ee
which are such that when a given component of the generalized Lie derivative is
projected by its corresponding intertwining tensor, the sub-algebra formed by
it, together with the previous components, closes. The reason for this is that
the contraction between successive intertwining tensors vanishes
\be
F_{\cal A}{}^{\alpha} F_{\alpha}{}^A = F_{\bf A}{}^{\cal A} F_{\cal
A}{}^{\alpha} = \dots = 0
\ee
and the failure of closure when the hierarchy is truncated to a given level, is
proportional to the intertwiner at that level.

A field-dependent generalized frame in the $E$-tangent space can be introduced
\be
\mathbb{E}_{\bar{\mathbb{A}}}{}^{\mathbb{A}} = \left(\begin{matrix}
e_{\bar{a}}{}^{\mu}
&-e_{\bar{a}}{}^{\rho}A_{\rho}{}^A &
- e_{\bar{a}}{}^{\rho}(B_{\rho\mu}{}^\alpha -
A_\rho{}^BA_\mu{}^C(t^{\alpha})_{BC}) & \mathbb{E}_{\bar a \nu \rho}{}^{\cal
A} \\
0 & \Phi_{\bar{A}}{}^M & -2 A_\mu{}^B \Phi_{\bar{A}}{}^C
(t^{\alpha})_{BC}&\dots \\
0 & 0 & - (e^{-1})_{\mu}{}^{\bar{a}}  (t_{\bar \alpha})^{\bar A \bar B}
\Phi_{\bar{A}}{}^A\Phi_{\bar{B}}{}^B (t^{\alpha})_{AB} & \dots \\
   &   &   & & \ddots
\end{matrix}\right)
\ee
where, in particular
\bea \label{genbein14}
\mathbb{E}_{\bar a \nu \rho}{}^{\cal A} &=& - e_{\bar a}{}^\mu
\left[\vphantom{\frac 1 3}C_{\mu\nu\rho}{}^{\cal A} + S^{\cal A}{}_{A\alpha}
A_\mu{}^A B_{\nu\rho}{}^\alpha \right.\\
&& \left. \ \ \ \ \ \ \ \ \ \ -\frac 1 3 S^{\cal A}{}_{A\alpha}
(t^{\alpha})_{BC} \left(A_\mu{}^A A_\nu{}^B A_\rho{}^C + 2 A_\mu{}^B A_\nu{}^C
A_\rho{}^A\right)\right]\nn
\eea
contains the $3$-form fields (and $S^{\cal A}{}_{A\alpha}$ is the projector from the $\bf 56 \times 133$ to the $\bf 912$ representation, given for example in \cite{west}), and the other components represented by the dots contain the remaining $p$-forms. This allows to make contact with the fields in gauged maximal supergravity, namely a
$4$-dimensional bein $e_{\bar a}{}^\mu$, scalars $\Phi_{\bar A}{}^A$,
gauge
vector fields $A_\mu{}^A$, and the (in)famous $p$-form fields that build the
tensor hierarchy $B_{\mu\nu}{}^\alpha$, $C_{\mu\nu\rho}{}^{\cal A}, \dots$.
Inserting this in the generalized
Lie derivative, we obtain the gauge transformations of each field
\bea
  \hat {\delta}_{  \xi} e_{\bar a}{}^{\mu} &=& L_\xi e_{\bar a}{}^\mu  \\
  \hat {\delta}_{  \xi} A_\mu{}^A &=& L_\xi A_\mu{}^A + \partial_\mu \xi^A +
F_{BC}{}^A \xi^B A_\mu{}^C - \xi_\mu{}^A \nn\\
  \hat {\delta}_{  \xi} \Phi_{\bar A}{}^A &=& L_\xi \Phi_{\bar A}{}^A +
F_{BC}{}^A \xi^B \Phi_{\bar A}{}^C \nn\\
  \hat {\delta}_{  \xi} B_{\mu \nu}{}^\alpha &=& L_\xi B_{\mu\nu}{}^\alpha  + 2
\partial_{[\mu} \xi_{\nu]}{}^\alpha  - \xi_{\mu\nu}{}^\alpha  + 2
(t^\alpha)_{BC} (A_{[\mu}{}^B \xi_{\nu]}{}^C - A_{[\mu}{}^B
\partial_{\nu]}\xi^C)\nn\\
  && - F_{A\beta}{}^\alpha \xi^A B_{\mu\nu}{}^\beta - 2 (t^\alpha)_{BC}
\xi^BF_\beta{}^C B_{\mu\nu}{}^\beta \nn\\ &\vdots& \nn
\eea
(where $L_{\xi}$ is the ordinary 4-dimensional Lie derivative along
$\xi^{\mu}$) which faithfully reproduce those of gauged maximal supergravity.

When the generalized Lie derivative is evaluated on frames, it defines the
generalized fluxes
\be
\mathbb{F}_{ \bar{\mathbb{A}}  \bar{\mathbb{B}}}{}^{ \bar{\mathbb{C}}} = (\hat
{\cal L}_{ \mathbb{E}_{\bar{\mathbb{A}}}} \mathbb{
E}_{\bar{\mathbb{B}}})^\mathbb{C} ( \mathbb{E}^{-1})_{{\mathbb{C}}}{}^{\bar
{\mathbb{C}}}
\ee
whose components determine the covariant quantities of gauged maximal
supergravity. We can list some of them 
\bea
\mathbb{F}_{\bar a \bar b}{}^{\bar c} &=& 2 e_{[\bar a}{}^\rho \partial_\rho
e_{\bar b]}{}^\sigma e_\sigma{}^{\bar c} = \omega_{[\bar a \bar b]}{}^{\bar c}\\
\mathbb{F}_{\bar a \bar b}{}^{\bar C} &=& - e_{\bar a}{}^\mu e_{\bar b}{}^\nu
(\Phi^{-1})_C{}^{\bar C}\ {\cal F}_{\mu\nu}{}^C\nn\\
\mathbb{F}_{\bar a \bar b \bar c}{}^{\bar \gamma} &=& e_{\bar a}{}^\mu e_{\bar
b}{}^\nu e_{\bar c}{}^\rho (t_\alpha)^{AB} (\Phi^{-1})_{A}{}^{\bar A}
(\Phi^{-1})_{B}{}^{\bar B} (t^{\bar \gamma})_{\bar A \bar B} {\cal
H}_{\mu\nu\rho}{}^\alpha \nn\\
\mathbb{F}_{\bar A \bar B}{}^{ \bar C} &=& \Phi_{\bar A}{}^A \Phi_{\bar B}{}^B
 (\Phi^{-1})_C{}^{\bar C} \ F_{AB}{}^C \nn\\
 \mathbb{F}_{\bar a \bar B}{}^{\bar C} & = & - \mathbb{F}_{\bar B \bar
a}{}^{\bar
C} = e_{\bar a}{}^\mu (\Phi^{-1})_C{}^{\bar C} \ D_\mu \Phi_{\bar B}{}^C\nn
 \eea
where
\bea
D_\mu \Phi_{\bar B}{}^C &=& \partial_\mu \Phi_{\bar B}{}^C - F_{AB}{}^C
A_\mu{}^A \Phi_{\bar B}{}^B\\
{\cal F}_{\mu\nu}{}^C &=& 2 \partial_{[\mu} A_{\nu]}{}^C - F_{[AB]}{}^C
A_\mu{}^A A_\nu{}^B + B_{\mu\nu}{}^\alpha F_\alpha{}^C\nn\\
{\cal H}_{\mu\nu\rho}{}^\alpha &=& 3 \left[\partial_{[\mu}
B_{\nu\rho]}{}^\alpha - C_{\mu\nu\rho}{}^{\cal A} F_{\cal A}{}^\alpha + 2
(t^\alpha)_{BC}\left(A_{[\mu}{}^B \partial_\nu A_{\rho]}{}^C + A_{[\mu}{}^B
B_{\nu\rho]}{}^\beta F_\beta{}^C \vphantom {\frac 1 3}\right.\right.\nn\\
&&\left.\left. \ \ \ \ \ \ \ \ \ \ \ \ \ \ \ \ \ \ \ \ \ \ \ \ \ \ \ \ \ \ \ \
\ \ \ \ \ \ \ \  \ \ \ \ \ \ \ \ \  + \ \frac 1 3\  F_{DE}{}^B A_{[\mu}{}^D
A_\nu{}^E A_{\rho]}{}^C\right)\right]\nn
\eea
Notice that the internal generalized fluxes that encode the gaugings
(\ref{gaugingsF}) arise here as particular components. Other components are the
antisymmetric part of the $4$-dimensional spin
connection, the curvatures of the $1$ and $2$-forms, the covariant derivatives
of the scalars, etc.

Since the generalized Lie derivative forms a closed algebra, and the
generalized fluxes are defined in terms of it, the closure conditions correspond
to Bianchi Identities (BI)
\be
\hat \Delta_{\bar{\mathbb{A}} \bar{\mathbb{B}}
\bar{\mathbb{C}}}{}^{\bar{\mathbb{D}}}= \left([\hat {\cal
L}_{\mathbb{E}_{\bar{\mathbb{A}}}}, \hat {\cal
L}_{\mathbb{E}_{\bar{\mathbb{B}}}}] \mathbb{E}_{\bar{\mathbb{C}}} - \hat {\cal
L}_{\hat {\cal L}_{\mathbb{E}_{\bar{\mathbb{A}}}}
\mathbb{E}_{\bar{\mathbb{B}}}}\mathbb{E}_{\bar{\mathbb{C}}}
\right){}^\mathbb{D}  (\mathbb{E}^{-1} )_{{\mathbb{D}}}{}^{\bar{\mathbb{D}}}= 0
\ee
These generalized BI include as particular components those of the four
dimensional Riemann tensor and
that of the curvature for the two form (for simplicity we compute only the
projection of the latter via the corresponding intertwiner)
\bea
\hat \Delta_{\bar d\bar a\bar b}{}^{\bar c} &=& 3 e_{\bar a}{}^\mu e_{\bar
d}{}^\nu e_{\bar b}{}^\rho (e^{-1})_\sigma{}^{\bar c} R_{[\mu\nu\rho]}{}^\sigma
= 3 (\partial_{[\bar a} \omega_{\bar d\bar b]}{}^{\bar c} + \omega_{[\bar a\bar
d}{}^{\bar e} \omega_{\bar b]\bar e}{}^{\bar c})
\\
\hat \Delta_{\bar d\bar a\bar b}{}^{\bar C} &=& e_{\bar d}{}^\mu e_{\bar
a}{}^\nu e_{\bar b}{}^\rho (\Phi^{-1})_M{}^{\bar C}\ (3 D_{[\mu} {\cal
F}_{\nu\rho]}{}^M - {\cal H}_{\mu\nu\rho}{}^M)
\eea
We note that they contain the BIs of General Relativity and those of the gauge
sector of maximal supergravity in a unified way.

\subsubsection{Generalized diffeomorphisms in extended geometry}

Next we explore the decompactified case, towards the construction of a full
generalized Lie derivative containing derivatives with respect to the external
and internal components. The generalized Lie derivative in the four-dimensional
case (\ref{reducedgeneralform}) takes the form of a ``gauged'' generalized Lie
derivative. Its structure coincides with that of Gauged DFT \cite{Hohm:2011ex},
which can be obtained from a higher-dimensional generalized Lie derivative
through a generalized Scherk-Schwarz reduction \cite{Grana:2012rr}. Here we
explore to what extent the gauged generalized Lie derivative
(\ref{reducedgeneralform}) admits an uplift to an extended space. However, we
will be general and not specify a particular U-duality group. Now, the
generalized vector fields also span an extended tangent space
\be
\hat \xi^{\mathbb{M}} = (\hat \xi^\mu, \hat \xi^{M}, \hat \xi_\mu{}^{<MN>}, \hat
\xi_{\mu\nu}{}^{<MNP>},\dots)
\ee
and moreover, we have put a hat on them to signal that they depend on both
external and internal coordinates, and recall the notation $<\dots>$ means a
projection to the relevant representations in the tensor hierarchy for the
different levels, where we name the level $p \geq 1$ as that of the $p-1$-form
gauge parameter.

 Schematically, the generalized Lie derivative adopts the following form
\be
\left(\hat {\cal L}_{\hat \xi_1} \hat \xi_2\right){}^{\mathbb{M}} = \hat
\xi_1^{\mathbb{P}} \partial_{\mathbb{P}} \hat \xi_2^{\mathbb{M}} - \hat
\xi_2^{\mathbb{P}} \partial_{\mathbb{P}} \hat \xi_1^{\mathbb{M}} +
Y^{\mathbb{M}}{}_{\mathbb{P}}{}^{\mathbb{N}}{}_{\mathbb{Q}}
\partial_{\mathbb{N}} \hat \xi_1^{\mathbb{Q}}\hat \xi_2^{\mathbb{P}}
 \ee
where derivatives now act with respect to internal directions also
$\partial_{\mathbb{M}} = (\partial_\mu, \partial_M,0,\dots)$. One could also
consider derivatives with respect to the extended tangent directions associated
to the $p$-forms, and
constrain them through a generalized section condition, but this is not the
approach we adopt here. The generalized $Y$-tensor contains $GL(d)$ and
$E_{n+1(n+1)}$ invariants, and when it is restricted to the internal sector it
coincides with the $Y$-tensors of the different U-duality groups, in particular
with (\ref{Ytensor}) for $E_{7(7)}$. The purely external components of it
vanish, and then one recovers the external diffeomorphisms of Riemannian
geometry. The details will be presented in Section \ref{sec:extendedst}. The
generalized $Y$-tensor is the one that projects the components of the
generalized vectors to
the relevant representation according to the U-duality group and the level of
the hierarchy.

Performing a generalized Scherk-Schwarz reduction $\hat \xi^{\mathbb{M}} =
U_{\mathbb{A}}{}^{\mathbb{M}}(Y) \xi^{\mathbb{A}}(x)$, and plugging it above,
one can make contact with the gauged generalized Lie derivative
(\ref{reducedgeneralform}). When the internal derivatives $\partial_M$ hit the
twist matrix $U(Y)$, it forms gaugings, that are contained in the last term in
(\ref{reducedgeneralform}). On the other hand, when the external derivatives
$\partial_\mu$ hit the $x$-dependent part, this reproduces the first three
terms in (\ref{reducedgeneralform}). Then, the $W$-tensor there is the
generalized $Y$-tensor here.

We have worked the hierarchy up to the $2$-form level, and found a couple of
intriguing facts. To begin with, the first level component $(\hat {\cal L}_{\hat \xi_1}
\hat \xi_2)^M$
contains terms that are projected by the first ``intertwining''
operator\footnote{ Notice that when $\frac{1}{2}Y^M{}_P{}^N{}_Q\
\partial_N(\ldots)^{PQ}$ acts on
twist matrices it is proportional to the intertwining tensor, i.e.
$\frac{1}{2}Y^M{}_P{}^N{}_Q\ \partial_N(U_A{}^PU_B{}^Q)=F_{(AB)}{}^CU_C{}^M$.
This happens only when the $Y^M{}_P{}^N{}_Q$ is symmetric in the $PQ$ indices.}
\be
\frac{1}{2}Y^M{}_P{}^N{}_Q\ \partial_N\ (\ldots)^{PQ}
\ee
and closes up to section condition-like terms (that compactify to quadratic
constraints) and terms proportional to $Y^M{}_{[P}{}^N{}_{Q]}$. While this
vanishes when the U-duality group is $E_{n+1(n+1)}$ with $n<6$,  we find a
closure obstruction for $E_{7(7)}$ already at the first level of the
hierarchy\footnote{ In $E_{8(8)}$ the failure of closure appears already in the purely internal
sector, even before coupling it to space-time \cite{Berman:2012vc}.}.
Regarding the second level component $(\hat {\cal L}_{\hat \xi} \hat V)_\mu{}^{<MN>}$, it
includes terms that are projected by the second ``intertwining'' operator
\be
\tilde Y^{MNT}{}_{QRS}\ \partial_T \ , \ \ \ \ {\rm where} \ \ \ \ \tilde
Y^{MNT}{}_{QRS} = Y^M{}_P{}^N{}_Q Y^P{}_R{}^T{}_S - Y^M{}_R{}^N{}_S
\delta^T_Q
\ee
The closure of this component is proportional to terms that depend on $\tilde
Y^{MNT}{}_{(QRS)}$. While this vanishes for $n<5$, we now find an
obstruction for $E_{6(6)}$. It then appears to be a pattern affecting the
$E_{n+1(n+1)}$ duality groups at the level $7-n$ of the tensor hierarchy. A
discussion on this point can be found in Section \ref{sec:extendedst}.

Finally, we have introduced a field-dependent generalized frame, and from it defined the fluxes and extracted the gauge transformations of its components. And we have also connected the tensor hierarchy with M-theory brane charges, and shown how our results reproduce the gauge transformations of the fields in $11$-dimensional supergravity.

\section{Generalized diffeomorphisms in gauged maximal supergravity}
\label{sec:E}

Our construction will begin with the canonical generalized Lie derivative in a
Yang-Mills theory coupled to gravity after a Kaluza-Klein decomposition. Since
in our case of interest the gaugings are not strictly speaking structure
constants (in this section we focus on the $4$-dimensional $E_{7(7)}$ case),
i.e. are not antisymmetric, the original proposal will fail to close
 and  require an extension. We will go through this extension in a systematic
way, ending with a closed form of a generalized Lie derivative for gauged maximal
supergravity. This analysis will serve as a base for the next extension,
pointing towards a full generalized Lie derivative in the mega-space-time with
manifest U-duality covariance.

\subsection{Generalized Lie derivative and closure}

We begin with a generalized Lie derivative, with the following components
\bea
( \hat {\cal L}_{ \xi_1}  \xi_2)^\mu &=& (L_{\xi_1} \xi_2)^\mu
\label{firstgenlie}\\
( \hat {\cal L}_{ \xi_1}  \xi_2)^A &=& L_{\xi_1}\xi_2^A -
\xi_2^\rho \partial_\rho\xi_1^A + F_{BC}{}^A \xi_1^B \xi_2^C \nn
\eea
where $L_{\xi_1}$ generate $4$-dimensional diffeomorphisms $\mu  = 1
,\dots 4$, and the extra components generate gauge transformations (with gauge
parameters $\xi^A$) $A = 1,
\dots, 56$. The constant gaugings $F_{AB}{}^C$ belong to the $\bf 56 + 912$
representations allowed by the linear supersymmetric constraint in gauged
maximal supergravity (see Section \ref{sec:summary}).

 Let us briefly state what the closure condition is. We are using the
convention that the generalized Lie derivative $\hat  {\cal L}$ acts on objects
assuming that they are generalized tensors. We can also define a generalized
gauge transformation $\hat  \delta$ that transforms objects without assuming
any covariancy properties. Let us consider an example to understand the
difference. While the generalized Lie derivative $\hat {\cal L}$ treats
$\partial_{\mathbb{A}} V^{\mathbb{B}}$ as if it were a tensor (we emphasize
that this is mere notation, since the generalized Lie derivative is only defined
to act on tensors), the gauge transformation $\hat \delta$ commutes with the
derivative, transforming this as $\hat \delta_\xi (\partial_{\mathbb{A}}
V^{\mathbb{B}})= \partial_{\mathbb{A}} ( \hat \delta_{ \xi}  V)^{\mathbb{B}} =
\partial_{\mathbb{A}} (\hat {\cal L}_{\xi}  V)^{\mathbb{B}}$. Then, we can
define the operator
\be
\hat \Delta_{ \xi} = \hat \delta_{\xi} - \hat {\cal L}_{ \xi}
\ee
which measures the failure of the covariance of the object on which it acts.
So, for example, we have that on vectors $\hat \Delta_{\xi}  V = 0$. In
particular, we would want the generalized Lie derivative to transform vectors
into vectors, this is the requirement known as closure constraint
\be
\left(\hat \Delta_{ {\xi}_1}  \hat {\cal L}_{{\xi}_2}  V\right){}^{\mathbb{A}} =
\left[\left(\left[\hat {\cal L}_{\xi_{1}},\hat {\cal L}_{\xi_{2}}\right]
-\hat {\cal L}_{ \hat {\cal L}_{\xi_1}\xi_2}\right) V\right]{}^{\mathbb{A}}= 0
\ee

Then, for the generalized Lie derivative (\ref{firstgenlie}), the closure
conditions become
\bea
(\hat \Delta_{\xi_1} \hat {\cal L}_{\xi_2} V )^\mu &=& 0 \label{falla1}\\
(\hat \Delta_{\xi_1} \hat {\cal L}_{\xi_2} V )^A &=& - 2 V^\rho F_{(BC)}{}^A
\partial_\rho\xi_1^B \xi_2^C + \left( [F_B, F_C] + F_{BC}{}^E
F_E\right){}_D{}^A \xi_1^B \xi_2^C V^D\nn
\eea
Since we are assuming that the gaugings satisfy the quadratic constraints, the
last term vanishes, and the failure of the closure is proportional to the
symmetric components of the gaugings $F_{(BC)}{}^A$. Symmetrized in this way,
the indices $(BC)$ belong to the adjoint $\bf 133$ representation of
$E_{7(7)}$, and $F_{(BC)}{}^A$ is called the intertwining tensor.

We then see that in order to achieve closure, the original generalized Lie
derivative has to be extended. Since the failure of closure is proportional to
$V^\mu$, we can add an additional term
\bea
(\hat {\cal L}_{ \xi_1}  \xi_2)^\mu &=& (L_{\xi_1} \xi_2)^\mu
\label{extraterm}\\
(\hat {\cal L}_{ \xi_1}  \xi_2)^A &=& L_{\xi_1}\xi_2^A -
\xi_2^\rho \partial_\rho\xi_1^A + F_{BC}{}^A \xi_1^B \xi_2^C +
\underline{\xi_2^\rho \xi_{1\rho}{}^A}\nn
\eea
containing a new gauge parameter $\xi_\rho{}^A$. Its transformation should be
such that it cancels the failure of the closure. A quick computation shows that
now (we impose the quadratic constraints on the gaugings)
\bea
(\hat \Delta_{\xi_1} \hat {\cal L}_{\xi_2} V )^\mu &=& 0 \\\nn
(\hat \Delta_{\xi_1} \hat {\cal L}_{\xi_2} V )^A &=&
V^\rho\big[(\delta_{\xi_1}\xi_2)_\rho{}^A
-(L_{\xi_1}\xi_2)_{\rho}{}^A-2\xi_2{}^{\sigma}\partial_{[\sigma}
\xi_{1\rho]}{}^A- 2 F_{(BC)}{}^A \partial_\rho
\xi_1^B \xi_2^C \\\nn
&& - 2 F_{BC}{}^A \xi_{[1}^B \xi_{2]\rho}{}^C\big]  + F_{BC}{}^A \xi_2^\rho
\xi_{1\rho}{}^B V^C\nn
\eea
While the first block between brackets dictates how the new gauge parameters
must transform, notice that the last term cannot be absorbed in the brackets,
and its vanishing must be imposed as a constraint
\be
\xi_2^\rho \xi_{1\rho}{}^B  F_{BC}{}^A V^C = 0
\ee
Recalling the quadratic constraints we can rapidly find a solution to this
equation
\be \xi_\mu{}^A = F_{(BC)}{}^A \xi_\mu{}^{BC} = F_{(BC)}{}^A (t_\alpha)^{BC}
\xi_\mu{}^\alpha  = F_\alpha{}^A \xi_\mu{}^\alpha\label{projection1}
\ee
The last step is possible because the intertwining tensor $F_\alpha{}^A$ takes
values in the adjoint $\bf 133$ representation of $E_{7(7)}$.   Of course, we
could have guessed from the beginning that the completion of the original
generalized Lie derivative would include components of this form, because the
failure for its closure is proportional to the intertwining tensor.

Now, if we generalize the notion of a vector, extending it to include
$\xi_\mu{}^{\alpha}$ as new components in an extended tangent space,
we now find a closed algebra of the form
\bea
( \hat {\cal L}_{ \xi_1}  \xi_2)^\mu &=& (L_{\xi_1}\xi_2)^\mu\\\nn
( \hat {\cal L}_{ \xi_1}  \xi_2)^A &=& L_{\xi_1}\xi_2^A -
\xi_2^\rho \partial_\rho\xi_1^A + F_{BC}{}^A \xi_1^B \xi_2^C + \xi_2^\rho
\xi_{1\rho}{}^A \nn\\
( \hat {\cal L}_{ \xi_1}  \xi_2)_\mu{}^A &=&
(L_{\xi_1}\xi_{2})_{\mu}{}^A+2\xi_{2}^{\sigma}\partial_{[\sigma}
\xi_{1\rho]}{}^A  + 2 F_{(BC)}{}^A (2 \xi_{[1}^B \xi_{2]\mu}{}^C +
\xi_2^{B}\partial_\mu\xi_1^{C})\nn
\eea
Here the last component is projected by the intertwining tensor from the $\bf
133$ representation to the $\bf 56$ representation $\xi_\mu{}^A = F_\alpha{}^A
\xi_\mu{}^\alpha$ as in (\ref{projection1}).
  Removing this projection is the topic of the next subsection. By now we have
found a closed (projected) generalized Lie derivative, that is enough to
reproduce the gauge transformation of the bosonic sector of maximal gauged
supergravity in the formulation of \cite{de Wit:2007mt}. In fact, after some
algebra one finds
\be
\left(\hat \Delta_{\xi_1} \hat {\cal L}_{\xi_2} V\right){}_\mu{}^A = 0
\ee
and the full closure is guaranteed.

\subsection{The next step in the hierarchy ({\bf 912})}

It follows from (\ref{projection1}) that the last component of the generalized
vectors are projected, and thus so is the last component of the generalized Lie
derivative. The projection is due to the intertwining tensor
\be
F_{\alpha}{}^A = (t_{\alpha}){}^{BC} F_{(BC)}{}^A \label{inter1}
 \ee
 We can then factorize it, and determine the un-projected components up to
terms that vanish due to the projection
\bea
F_\alpha{}^A\ \ \left[\vphantom{\frac 12}( \hat {\cal L}_{ \xi_1}
\xi_2)_\mu{}^\alpha \right.&=&
(L_{\xi_1}\xi_{2})_{\mu}{}^{\alpha}+2\xi_{2}^{\sigma}\partial_{[\sigma}
\xi_{1\mu]}{}^{\alpha} \nn\\
&&\left.-\ 2 (t^{\alpha})_{BC} \left( 2 \xi_{[1}^B \xi_{2]\mu}{}^\beta
F_\beta{}^C  + \xi_2^B \partial_\mu \xi_1^C\right) +
\Gamma_\mu{}^\alpha\vphantom{\frac 12}\right]\label{projectedGamma}
\eea
Here $\Gamma_\mu{}^\alpha$ is the collection of terms that vanish due to the
projection, i.e. it satisfies $\Gamma_\mu{}^\alpha F_\alpha{}^A = 0$.
Setting for the moment $\Gamma_\mu{}^\alpha = 0$, we can compute closure of
this last un-projected component, finding
\bea
\left(\hat \Delta_{\xi_1}\hat {\cal L}_{\xi_2} \xi_3 \right){}_\mu{}^\alpha &=&
-
F_{ABC}{}^\alpha\left[(2 \xi_3^\rho\xi_{[1\rho}{}^\gamma \xi_{2]\mu}{}^\beta -
\xi_{3\mu}{}^\gamma \xi_2^\rho \xi_{1\rho}{}^\beta) F_\gamma{}^A
(t_{\beta})^{BC} \right.\nn\\
&& \ \ \ \ \ \ \ \ \ \ \ \left.  +2 \xi_{3\mu}{}^\gamma F_\gamma{}^C
\xi_1^A\xi_2^B+ 4 \xi_3^B (\xi_{[2}^A \xi_{1]\mu}{}^\gamma F_\gamma{}^C -
\xi_{[2}^A \partial_\mu \xi_{1]}^C)         \right]\label{unprojectedfailure}
\eea
Here we have used the quadratic constraints and defined
\be
F_{ABC}{}^\alpha = 2 (F_{A(B}{}^D (t^\alpha)_{C)D} - F_{(BC)}{}^D
(t^{\alpha})_{DA}) \label{inter2}
\ee
It is easy to see that this tensor satisfies the following properties
\be
P{}^A{}_B{}^C{}_D \ F_{EC}{}^{D\alpha} = F_{EB}{}^{A\alpha} \ , \ \ \ \
F_{(ABC)}{}^\alpha = 0 \  , \ \ \ \ F_{AB}{}^{B\alpha} = F_{BA}{}^{B\alpha} = 0
\ee
and then  its indices $A(BC)$ belong to the $\bf 912$ representation of
$E_{7(7)}$ \cite{de Wit:2007mt}.

Clearly, since the projected components enjoy a closed algebra, the failure of
the un-projected components must cancel through a projection with the
intertwining tensor. A short computation shows that
\be
F_{ABC}{}^\alpha F_\alpha{}^D = 0
\ee
due to the quadratic constraints. This makes clear that $\Gamma_{\mu}{}^\alpha$
should be proportional to this tensor. Also, it must be selected so as to
cancel the un-projected contributions (\ref{unprojectedfailure}). After some
algebra we find that the correction to the un-projected generalized Lie
derivative is given by
\be
\Gamma_\mu{}^\alpha = \xi_2^\rho \xi_{1\rho\mu}{}^\alpha - F_{A\beta}{}^\alpha
\xi_{2\mu}{}^\beta \xi_1^A\label{GammaMuAlpha}
\ee
Here, we have denoted the indices in the $\bf 912$ as $A\beta$, and introduced
$133$ new gauge two-form gauge parameters $ \xi_{\rho\mu}{}^\alpha$. However,
these are now projected by the new intertwining tensor $F_{A\beta}{}^\alpha$,
which projects the $\bf 912$ into the $\bf 133$, so this component of the
generalized Lie derivative only knows about the projection of the new gauge
parameters
 \be
 \xi_{\mu \nu}{}^\alpha = F_{A\beta}{}^\alpha \xi_{\mu
\nu}{}^{A\beta}\label{proj2}
 \ee
  This is analog to the previous intertwining $F_\alpha{}^A$, which projects
the $\bf 133$ into the $\bf 56$. Introducing (\ref{GammaMuAlpha}) in
(\ref{projectedGamma}), we find that closure is achieved provided the gauge
transformation of the (projection of the) new gauge parameters is given by
\bea
\left(\hat {\cal L}_{\xi_1}\xi_2\right){}_{\mu \nu}{}^{\alpha} &=&
(L_{\xi_1}\xi_2)_{\mu\nu}{}^{\alpha}-3\xi_2^{\rho}\partial_{[\rho}\xi_{1\mu\nu]}
{}^{\alpha}\\
&& + 2 F_{A\beta}{}^\alpha \left(\xi_{2[\mu}{}^\beta \partial_{\nu]}\xi_1^A  -
\xi_2^A \partial_{[\mu}\xi_{1\nu]}{}^\beta - \xi_{[1}^A \xi_{2]\mu\nu}{}^\beta
+ \xi_{1[\mu}{}^\beta \xi_{2\nu]}{}^\gamma F_\gamma{}^A\right)\nn
\eea

Then, the following algebra closes up to the quadratic constraints
\bea
\left( \hat {\cal L}_{ \xi_1} \xi_2\right)^\mu &=& (L_{\xi_1}\xi_2)^\mu
\label{SecondStep}\\
\left( \hat {\cal L}_{  \xi_1}   \xi_2\right)^A &=& L_{\xi_1}\xi_2^A -
\xi_2^\rho \partial_\rho\xi_1^A + F_{BC}{}^A \xi_1^B \xi_2^C + \xi_2^\rho
\xi_{1\rho}{}^\gamma F_\gamma{}^A \nn\\
\left(\hat {\cal L}_{ \xi_1}  \xi_2\right)_\mu{}^\alpha &=&
(L_{\xi_1}\xi_{2})_{\mu}{}^{\alpha} - 2\xi_{2}^{\rho}\partial_{[\rho}
\xi_{1\mu]}{}^{\alpha}-\ 2 (t^{\alpha})_{BC}
\left( 2 \xi_{[1}^B \xi_{2]\mu}{}^\gamma F_\gamma{}^C  + \xi_2^B \partial_\mu
\xi_1^C\right)\nn\\
&& +\ \xi_2^\rho \xi_{1\rho\mu}{}^\alpha - F_{A\beta}{}^\alpha
\xi_{2\mu}{}^\beta \xi^A_1\nn\\
\left(\hat {\cal L}_{\xi_1}\xi_2\right){}_{\mu \nu}{}^{\alpha} &=&
(L_{\xi_1}\xi_2)_{\mu\nu}^{\alpha}-3\xi_2^{\rho}\partial_{[\rho}\xi_{1\mu\nu]}{}
^{\alpha} \nn\\
&& + 2 F_{A\beta}{}^\alpha \left(\xi_{2[\mu}{}^\beta \partial_{\nu]}\xi_1^A  -
\xi_2^A \partial_{[\mu}\xi_{1\nu]}{}^\beta - \xi_{[1}^A \xi_{2]\mu\nu}{}^\beta
+ \xi_{1[\mu}{}^\beta \xi_{2\nu]}{}^A\right)\nn
\eea

\subsection{The next step in the hierarchy ({\bf 133 + 8645}) and so on...}

Recalling (\ref{proj2}) we see that the last component in (\ref{SecondStep}) is
actually the result of a new projection due to the new intertwining tensor
\bea
F_{\cal A}{}^\alpha&&\!\!\!\!\!\!\!\left[\vphantom{\frac 12}\left(\hat {\cal
L}_{\xi_1}\xi_2\right)_{\mu \nu}{}^{\cal A} \right.=
\left(L_{\xi_1}\xi_2\right)_{\mu \nu}{}^{\cal A} - 3 \xi_2^\rho
\partial_{[\rho} \xi_{1\mu\nu]}{}^{\cal A} \\
&& \left. \ \ \ \ \ \ \ \ \ \ + 2 S^{\cal A}{}_{B\delta}
\left(\xi_{2[\mu}{}^\delta \partial_{\nu]}\xi_1^B  -  \xi_2^B
\partial_{[\mu}\xi_{1\nu]}{}^\delta - \xi_{[1}^B \xi_{2]\mu\nu}{}^{\cal B}
F_{\cal B}{}^\delta - \xi_{2[\mu}{}^\delta \xi_{1\nu]}{}^\gamma
F_\gamma{}^B\right) + \Gamma_{\mu\nu}{}^{\cal A}\vphantom{\frac 12}\ \right]\nn
\eea
where we have introduced potential new contributions that vanish due to the
projection
\be
\Gamma_{\mu\nu}{}^{\cal A} F_{\cal A}{}^\alpha = 0
\ee
and defined $S^{\cal A}{}_{B\delta}$ as a projector to the $\bf 912$.

We can now proceed as in the previous subsection, and evaluate the failure of
the closure for the unprojected components, setting for the moment
$\Gamma_{\mu\nu}{}^{\cal A} = 0$. A long computation shows that
\bea
\left(\hat \Delta_{\xi_2} \hat {\cal L}_{\xi_1} \xi_2\right)_{\mu\nu}{}^{\cal
A} \!\!&=&\!\! F_{BC\alpha}{}^{\cal A} \left[\vphantom{\frac 1 2}3 \xi_2^\rho
\left((t_\beta)^{BC} \left(2\xi_{1[\mu}{}^\alpha\partial_\nu
\xi_{3\rho]}{}^\beta + (\xi_{3[\rho}{}^\alpha \xi_{1\mu\nu]}{}^{\cal B} -
\xi_{1[\rho}{}^\alpha\xi_{3\mu\nu]}{}^{\cal B})F_{\cal B}{}^\beta\right)
\right.\right.\nn\\
&& \ \ \  \left. - S_{\cal B}{}^{C\alpha} \left( \xi_{1[\mu\nu}{}^{\cal B}
\partial_{\rho]}\xi_3^B + \xi_1^B \partial_{[\rho} \xi_{3\mu\nu]}{}^{\cal
B}\right) \right) + (t_\beta)^{BC} \xi_1^\rho \xi_{3\rho}{}^\alpha
\xi_{2\mu\nu}{}^{\cal B} F_{\cal B}{}^\beta \\
&& \!\!\!\!\!\! + \left.2 \xi_{2[\mu}{}^\alpha \xi_{3\nu]}{}^\gamma
F_\gamma{}^C \xi_1^B - 2 \xi_{2[\mu}{}^\alpha \xi_{1\nu]}{}^\gamma F_\gamma{}^C
\xi_3^B + (2 \xi_{[1}^B \xi_{3]\mu\nu}{}^{\cal B} \xi_2^C + \xi_3^B \xi_1^C
\xi_{2\mu\nu}{}^{\cal B}) F_{\cal B}{}^\alpha \vphantom{\frac12}\right]\nn
\eea

As before, we have been able to factorize the new intertwining tensor
\bea
F_{BC\alpha}{}^{\cal A} = -2 S^{\cal A}{}_{C\beta} (t^\beta)_{BD} F_\alpha{}^D
+ S^{\cal A}{}_{D\alpha} F_{BC}{}^D + 2 S^{\cal A}{}_{[B| \beta}
F_{|C]\alpha}{}^\beta \label{inter3}
\eea
We will collectively denote its indices ${\bf A} = BC\alpha$. As expected, this
intertwining tensor is canceled through a projection with the previous one
\be
F_{\bf A}{}^{\cal A} F_{\cal A}{}^\alpha = 0
\ee
due to the quadratic constraints. It also satisfies the properties
\be
F_{BC}{}^{\alpha{\cal A}} = P_{(912)C}{}^\alpha,^D{}_\beta \
F_{BD}{}^{\beta{\cal A}} \ , \ \ \ \ \ \ \ F_{(BC)\alpha}{}^{\cal A} = 2
(t^\beta)_{BC} S^{\cal A}{}_{D[\beta} F_{\alpha]}{}^D
\ee
The indices $C\alpha$ actually belong to the $\bf 912$ and ${\bf A}=BC\alpha$
belongs to the $\bf 8645 + 133$.

Now, following the route of the previous section, and inspired by
(\ref{GammaMuAlpha}), we propose
\be
\Gamma_{\mu\nu}{}^{\cal A}  = \xi_2^\rho  \xi_{1\rho\mu\nu}{}^{\cal A} +
F_{A{\cal B}}{}^{\cal A} \xi_{2\mu\nu}{}^{\cal B} \xi_1^A
\ee
where we have introduced new gauge parameters $\xi_{\rho\mu\nu}{}^{\cal A} =
F_{\bf A}{}^{\cal A} \xi_{\rho\mu\nu}{}^{\bf A}$. Another long computation
shows that now full closure is achieved if
\bea
\left(\hat {\cal L}_{\xi_1}\xi_2\right){}_{\rho\mu\nu}{}^{\cal A} &=&
\left(L_{\xi_1}\xi_2\right)_{\rho\mu\nu}{}^{\cal A} - 4 \xi_2^\sigma
\partial_{[\sigma}\xi_{1\rho\mu\nu]}{}^{\cal A} \\\nn
&& + F_{A{\cal B}}{}^{\cal A} \left[ \vphantom{\frac 1 2} 3(t_{\beta})^{AC}
S^{\cal B}{}_{C\alpha} \left( 2 \xi_{2[\mu}{}^\alpha \partial_\nu
\xi_{1\rho]}{}^\beta + \xi_{1[\rho}{}^\alpha\xi_{2\mu\nu]}{}^\beta -
\xi_{2[\rho}{}^\alpha \xi_{1\mu\nu]}{}^\beta  \right)\right.\nn\\
&& \ \ \ \ \ \ \ \ \ \ \ \ \ + \left. \vphantom{\frac 1 2}  3
\xi_{2[\mu\nu}{}^{\cal B} \partial_{\rho]}\xi_1^A + 3 \xi_2^A \partial_{[\rho}
\xi_{1\mu\nu]}{}^{\cal B} - 2 \xi_{[2}^A \xi_{1]\rho\mu\nu}{}^{\cal B}\right]\nn
\eea

Again, one could now extract the projection of the intertwining tensor from
this expression, and add new terms to achieve closure, which will take the form
\be
\Gamma_{\rho\mu\nu}{}^{\bf A} = \xi_2^\sigma \xi_{1\sigma\rho\mu\nu}{}^{\bf A}
- F_{A{\bf B}}{}^{\bf A} \xi_{2\rho\mu\nu}{}^{\bf B} \xi_1^A
\ee
with $F_{A{\bf B}}{}^{\bf A}$ the next intertwining tensor. In principle one
can repeat these steps over and over and build the so-called tensor hierarchy.
Note however that a $4$-form gauge parameter is supposed to transform a
$5$-form, which cannot be present in $4$-dimensions, and we will then stop here.

\subsection{The full generalized Lie derivative}

We have been able to construct, step by step, a generalized Lie derivative that
incorporates the diffeomorphisms and gauge transformations of gauged maximal
supergravity. Schematically, it takes the form
\be
\left(\hat {\cal L}_{\xi_1} \xi_2\right)^{\mathbb{A}} =
\xi_1^{\mathbb{B}} \partial_{\mathbb{B}} \xi_2^{\mathbb{A}} -
\xi_2^{\mathbb{B}} \partial_{\mathbb{B}} \xi_1^{\mathbb{A}}  +
W^{\mathbb{A}}{}_{\mathbb{B}}{}^{\mathbb{C}}{}_{\mathbb{D}}\partial_{\mathbb{C}}
 \xi_1^{\mathbb{D}} \xi_2^{\mathbb{B}} + F_{\mathbb{BC}}{}^{\mathbb{A}}
\xi_1^{\mathbb{B}} \xi_2^{\mathbb{C}} \label{generalform}
\ee
where we have collectively denoted the indices $\xi^{\mathbb{A}} = (\xi^\mu,
\xi^A, \xi_{\mu}{}^\alpha,\xi_{\mu\nu}{}^{\cal A}, \xi_{\mu\nu\rho}{}^{\bf
A},\dots)$, and since we are working in $4$-dimensions we also have
$\partial_{\mathbb{A}} = (\partial_\mu, 0 ,\dots)$. The first three terms are
un-gauged, and the $F_{\mathbb{BC}}{}^{\mathbb{A}}$ represents a collection of
all the gaugings, mostly containing intertwiners.
The $W^{\mathbb{A}}{}_{\mathbb{B}}{}^{\mathbb{C}}{}_{\mathbb{D}}$ tensor is
formed by invariants of the symmetry group. Notice that the generalized Lie
derivative is linear in derivatives. In particular, the gauged terms are not
derived, but the gaugings are linear in internal derivatives. Also notice that
only the gauge parameters that generate the transformation are derived. Its
structure resembles the general structure of the gauged generalized Lie
derivative of gauged DFT \cite{Hohm:2011ex}, \cite{Grana:2012rr},
\cite{Berman:2013cli}. The difference here is that the hierarchy of vectors
requires a large extended (exceptional) tangent space, and this is due to the
fact that the gaugings are not antisymmetric, and then there is a tower of
intertwiners. The remarkable equivalence between the structure of both
generalized Lie derivatives however suggests that this construction can be
uplifted to a duality covariant construction in an extended space, as it is
case for gauged DFT \cite{Grana:2012rr}. Such a construction should be equipped
with an un-gauged generalized Lie derivative in an extended space, and should
reduce to this one upon generalized Scherk-Schwarz reduction. We will give
later the first steps in this direction. Moreover, gauged DFT encodes
half-maximal gauged supergravities, and we will see later that this construction
encodes the maximal gauged supergravities in $4$-dimensions.

In components, the generalized Lie derivative (\ref{generalform}) reads
\bea
\left( \hat {\cal L}_{ \xi_1} \xi_2\right)^\mu &=& (L_{\xi_1}\xi_2)^\mu
\label{componentsgeneralform}\\
\left( \hat {\cal L}_{  \xi_1}   \xi_2\right)^A &=& (L_{\xi_1}\xi_2)^A -
\xi_2^\rho
\partial_\rho\xi_1^A  +  F_{[BC]}{}^A \xi_1^B \xi_2^C\nn\\
&&+\ F_{(BC)}{}^A \xi_2^B \xi_1^C + \xi_2^\rho \xi_{1\rho}{}^A \nn\\
\left(\hat {\cal L}_{ \xi_1}  \xi_2\right)_\mu{}^\alpha &=&
(L_{\xi_1}\xi_2)_\mu{}^\alpha
- 2\xi_2^\rho \partial_{[\rho}\xi_{1\mu]}{}^\alpha-  S^\alpha{}_{CB}
\left(2\xi_2^C \partial_\mu \xi_1^B + 4 \xi_{[1}^C \xi_{2]\mu}{}^\beta
F_\beta{}^B  \right)\nn\\
&& -\ F_{A\beta}{}^\alpha \xi_{2\mu}{}^\beta \xi^A_1+\ \xi_2^\rho
\xi_{1\rho\mu}{}^\alpha\nn\\
\left(\hat {\cal L}_{\xi_1}\xi_2\right){}_{\mu \nu}{}^{\cal A} &=&
\left(L_{\xi_1}\xi_2\right)_{\mu \nu}{}^{\cal A} - 3 \xi_2^\rho
\partial_{[\rho} \xi_{1\mu\nu]}{}^{\cal A}\nn\\
&&  -  S^{\cal A}{}_{C\beta} \left(2  \xi_2^C
\partial_{[\mu}\xi_{1\nu]}{}^\beta-2\xi_{2[\mu}{}^\beta \partial_{\nu]}\xi_1^C
 +2 \xi_{[1}^C \xi_{2]\mu\nu}{}^{\cal B} F_{\cal B}{}^\beta +2
\xi_{2[\mu}{}^\beta \xi_{1\nu]}{}^\gamma F_\gamma{}^C\right)\nn\\
 && +\ F_{A{\cal B}}{}^{\cal A} \xi_{2\mu\nu}{}^{\cal B} \xi_1^A + \xi_2^\rho
\xi_{1\rho\mu\nu}{}^{\cal A}\nn\\
 \left(\hat {\cal L}_{\xi_1}\xi_2\right){}_{\rho\mu\nu}{}^{\bf  A} &=&
\left(L_{\xi_1}\xi_2\right)_{\rho\mu\nu}{}^{\bf  A} - 4 \xi_2^\sigma
\partial_{[\sigma}\xi_{1\rho\mu\nu]}{}^{\bf A}\nn\\
&& - S^{\bf A}{}_{C{\cal B}} \left[ \vphantom{\frac 1 2} 3(t_{\beta})^{CD}
S^{\cal B}{}_{D\alpha} \left( 2 \xi_{2[\mu}{}^\alpha \partial_\nu
\xi_{1\rho]}{}^\beta + (\xi_{1[\rho}{}^\alpha\xi_{2\mu\nu]}{}^{\cal D} -
\xi_{2[\rho}{}^\alpha \xi_{1\mu\nu]}{}^{\cal D}) F_{\cal D}{}^\beta
\right)\right.\nn\\
&& \ \ \ \ \ \ \ \ \ \ \ \ \ + \left. \vphantom{\frac 1 2}  3 \xi_2^C
\partial_{[\rho} \xi_{1\mu\nu]}{}^{\cal B}+3 \xi_{2[\mu\nu}{}^{\cal B}
\partial_{\rho]}\xi_1^C +  2 \xi_{[1}^C \xi_{2]\rho\mu\nu}{}^{\bf B} F_{\bf
B}{}^{\cal B}\right]\nn\\
&& -\ F_{A{\bf B}}{}^{\bf A} \xi_{2\rho\mu\nu}{}^{\bf B} \xi_1^A + \xi_2^\sigma
\xi_{1\sigma\rho\mu\nu}{}^{\bf A} \nn\\
&\vdots& \nn
\eea
Here, the $S$-tensors correspond to projectors to the different
representations. One can rapidly identify a common structure in all the
components. There is a hierarchy of intertwining tensors
\be \label{intertwining}
F_{\alpha}{}^A\ , \ \ \ F_{\cal A}{}^{\alpha} \ , \ \ \ F_{\bf A}{}^{\cal A} \
, \ \ \ \dots
\ee
which have been defined in (\ref{inter1}), (\ref{inter2}) and (\ref{inter3}).
They are such that when a given component of the generalized Lie derivative is
projected by its corresponding intertwining tensor, the sub algebra formed by
it, together with the previous components, closes. The reason for this is that
the contraction between successive intertwining tensors vanishes
\be
F_{\cal A}{}^{\alpha} F_{\alpha}{}^A = F_{\bf A}{}^{\cal A} F_{\cal
A}{}^{\alpha} = \dots = 0
\ee
and the failure of closure when the hierarchy is truncated to a given level, is
proportional to the intertwiner of that level.

Formally, one can extend this into an infinite hierarchy \cite{Palmkvist:2013vya}, but beyond the dimension of space-time the fields would vanish, and then the levels considered here are the physically relevant ones.

\subsection{Generalized bein}

We now have a closed form of the generalized Lie derivative. Due to partial
projections by the intertwining tensors $F_\alpha{}^A$, $F_{\cal A}{}^\alpha$,
$F_{\bf A}{}^{\cal A}$, $\dots$ one can achieve closure step by step. Here, we
will propose a parameterization of the generalized frame or bein in terms of
the bosonic degrees of freedom of gauged maximal supergravity, namely
\begin{itemize}
\item A $4$-dimensional vielbein $e_{\bar a}{}^\mu$, where the flat index $\bar
a = 1,\dots, 4$ is acted on by the Lorentz group $SO(1,3)$.
\item $56$ gauge vector fields $A_\mu{}^A$.
\item $70$ scalars, parameterized by the coset matrix $\Phi_{\bar A}{}^A$,
where the flat index $\bar A = 1,\dots,56$ is acted on by $SU(8)$.
\item $133$ two-forms $B_{\mu\nu}{}^\alpha$, which at the level of the
Lagrangian are projected to $56$ components due to the intertwining tensor
$F_\alpha{}^A$ in the formulation of \cite{de Wit:2007mt}.
\item $912$ three-forms $C_{\mu\nu\rho}{}^{\cal A}$, with no dynamical degrees
of freedom.
\end{itemize}
Moreover, enlarging the generalized frame to the full E-tangent space, we could add more and more fields in the tensor
hierarchy, but we will stop here for simplicity.

Introducing
indices $\mathbb{A} = (^\mu,^A,_\mu{}^\alpha,_{\mu\nu}{}^{\cal
A},_{\mu\nu\rho}{}^{\bf A}, \dots)$ and $\bar {\mathbb{A}} = (^{\bar a},^{\bar
A},_{\bar a}{}^{\bar \alpha},_{\bar a\bar b}{}^{\bar{\cal A}},_{\bar a\bar b\bar
c}{}^{\bar{\bf A}}, \dots)$, we propose
\be
\mathbb{E}_{\bar{\mathbb{A}}}{}^{\mathbb{A}} = \left(\begin{matrix}
e_{\bar{a}}{}^{\mu}
&-e_{\bar{a}}{}^{\rho}A_{\rho}{}^A &
- e_{\bar{a}}{}^{\rho}(B_{\rho\mu}{}^\alpha -
A_\rho{}^BA_\mu{}^C(t^{\alpha})_{BC}) & \mathbb{E}_{\bar a \nu \rho}{}^{\cal
A} \\
0 & \Phi_{\bar{A}}{}^M & -2 A_\mu{}^B \Phi_{\bar{A}}{}^C
(t^{\alpha})_{BC}&\dots \\
0 & 0 & - (e^{-1})_{\mu}{}^{\bar{a}}  (t_{\bar \alpha})^{\bar A \bar B}
\Phi_{\bar{A}}{}^A\Phi_{\bar{B}}{}^B (t^{\alpha})_{AB} & \dots \\
   &   &   & & \ddots
\end{matrix}\right) \label{GenBein1}
\ee
where the flat indices are naturally acted on by $H = SO(1,3) \times SU(8)$,
$\mathbb{E}_{\bar a \nu \rho}{}^{\cal A}$ was defined in (\ref{genbein14}), and the remaining components would contain the higher $p$-forms.

Inserting this in the generalized Lie derivative, we obtain the gauge
transformations of each component
\bea
  \hat {\delta}_{  \xi} e_{\bar a}{}^{\mu} &=& L_\xi e_{\bar a}{}^\mu \label{componentsgaugetransf}\\
  \hat {\delta}_{  \xi} A_\mu{}^A &=& L_\xi A_\mu{}^A + \partial_\mu \xi^A +
F_{BC}{}^A \xi^B A_\mu{}^C - \xi_\mu{}^A \nn\\
  \hat {\delta}_{  \xi} \Phi_{\bar A}{}^A &=& L_\xi \Phi_{\bar A}{}^A +
F_{BC}{}^A \xi^B \Phi_{\bar A}{}^C \nn\\
  \hat {\delta}_{  \xi} B_{\mu \nu}{}^\alpha &=& L_\xi B_{\mu\nu}{}^\alpha  + 2
\partial_{[\mu} \xi_{\nu]}{}^\alpha  - \xi_{\mu\nu}{}^\alpha  + 2
(t^\alpha)_{BC} (A_{[\mu}{}^B \xi_{\nu]}{}^C - A_{[\mu}{}^B
\partial_{\nu]}\xi^C)\nn\\
  && - F_{A\beta}{}^\alpha \xi^A B_{\mu\nu}{}^\beta - 2 (t^\alpha)_{BC}
\xi^BF_\beta{}^C B_{\mu\nu}{}^\beta\nn \\
&\vdots& \nn
\eea
which faithfully reproduce those of gauged maximal supergravity in the
formulation of \cite{de Wit:2007mt}. Considering the remaining components would allow to make contact with \cite{west}.

Following the usual geometric
constructions in extended geometries, building generalized connections and
curvatures, one should be able to reproduce the bosonic sector of gauged
maximal supergravity in \cite{de Wit:2007mt} and even construct a democratic formulation containing the other $p$-forms. In fact, we will see in the next
subsection that the so-called generalized fluxes encode all the covariant
structures of the theory.

\subsection{Generalized fluxes}
We now define the so-called generalized fluxes
\be
\mathbb{F}_{ \bar{\mathbb{A}}  \bar{\mathbb{B}}}{}^{ \bar{\mathbb{C}}} = (\hat
{\cal L}_{ \mathbb{E}_{\bar{\mathbb{A}}}} \mathbb{
E}_{\bar{\mathbb{B}}})^\mathbb{C} (\mathbb{E}^{-1})_{{\mathbb{C}}}{}^{
\bar{\mathbb{C}}} \label{genfluxes}
\ee
Since they are defined through the generalized Lie derivative, they can only
define quantities that are covariant with respect to the gauge transformations.
Moreover, since the vectors involved in their definition are given by
generalized beins, we expect them to correspond to covariant derivatives and
curvatures. The simplest ones that depend on the degrees of freedom in
(\ref{GenBein1}) are
\bea
\mathbb{F}_{\bar a \bar b}{}^{\bar c} &=& 2 e_{[\bar a}{}^\rho \partial_\rho
e_{\bar b]}{}^\sigma e_\sigma{}^{\bar c} = \omega_{[\bar a \bar b]}{}^{\bar c}
\label{genfluxcomponents}\\
\mathbb{F}_{\bar a \bar b}{}^{\bar C} &=& - e_{\bar a}{}^\mu e_{\bar b}{}^\nu
(\Phi^{-1})_C{}^{\bar C}\ {\cal F}_{\mu\nu}{}^C\nn\\
\mathbb{F}_{\bar a \bar b \bar c}{}^{\bar \gamma} &=& e_{\bar a}{}^\mu e_{\bar
b}{}^\nu e_{\bar c}{}^\rho (t_\alpha)^{AB} (\Phi^{-1})_{A}{}^{\bar A}
(\Phi^{-1})_{B}{}^{\bar B} (t^{\bar \gamma})_{\bar A \bar B} {\cal
H}_{\mu\nu\rho}{}^\alpha \nn\\
\mathbb{F}_{\bar A \bar B}{}^{ \bar C} &=& \Phi_{\bar A}{}^A \Phi_{\bar B}{}^B
 (\Phi^{-1})_C{}^{\bar C} \ F_{AB}{}^C \nn\\
 \mathbb{F}_{\bar a \bar B}{}^{\bar C} &=& - \mathbb{F}_{\bar B \bar a}{}^{\bar
C} = e_{\bar a}{}^\mu (\Phi^{-1})_C{}^{\bar C} \ D_\mu \Phi_{\bar B}{}^C\nn
 \eea
where
\bea
D_\mu \Phi_{\bar B}{}^C &=& \partial_\mu \Phi_{\bar B}{}^C - F_{AB}{}^C
A_\mu{}^A \Phi_{\bar B}{}^B\\
{\cal F}_{\mu\nu}{}^C &=& 2 \partial_{[\mu} A_{\nu]}{}^C - F_{[AB]}{}^C
A_\mu{}^B A_\nu{}^C + B_{\mu\nu}{}^\alpha F_\alpha{}^C\nn\\
{\cal H}_{\mu\nu\rho}{}^\alpha &=& 3 \left[\partial_{[\mu}
B_{\nu\rho]}{}^\alpha - C_{\mu\nu\rho}{}^{\cal A} F_{\cal A}{}^\alpha + 2
(t^\alpha)_{BC}\left(A_{[\mu}{}^B \partial_\nu A_{\rho]}{}^C + A_{[\mu}{}^B
B_{\nu\rho]}{}^\beta F_\beta{}^C \vphantom {\frac 1 3}\right.\right.\nn\\
&&\left.\left. \ \ \ \ \ \ \ \ \ \ \ \ \ \ \ \ \ \ \ \ \ \ \ \ \ \ \ \ \ \ \ \
\ \ \ \ \ \ \ \  \ \ \ \ \ \ \ \ \  + \ \frac 1 3\  F_{DE}{}^B A_{[\mu}{}^D
A_\nu{}^E A_{\rho]}{}^C\right)\right]\nn
\eea
Here we can rapidly identify the antisymmetric part of the $4$-dimensional spin
connection, the curvatures of the $1$-forms and $2$-forms, the covariant
derivatives of the scalars and the gaugings. All these correspond to covariant
quantities. Note that to compute the curvature of the two-form, we need the
bein component defied
in (\ref{genbein14}). As before, one can further extend this analysis so as to
make contact with the higher-level curvatures
in \cite{west}.

In extended geometric constructions, the action (or generalized Ricci scalar)
is quadratic in generalized fluxes (see for example \cite{Grana:2012rr}). Then,
the covariant generalized Ricci scalar constructed from these generalized
fluxes will include a $4$-dimensional Ricci scalar originated from the spin
connection above, kinetic terms for the scalars originated from the covariant
derivatives of scalars above, kinetic terms for the gauge fields originated from
the field strengths above, and so on. In other words, the generalized fluxes we
have found are precisely the covariant quantities that enter the action of
gauged maximal supergravity. Furthermore, the other fluxes contained in this
formulation would allow to build in a generalized geometrical fashion a
democratic formulation of $4$-dimensional maximal supergravity, like the one
explored in \cite{Bergshoeff:2009ph}.

\subsection{Generalized Bianchi Identities}

Given that the generalized Lie derivative transforms tensors into tensors, the
closure conditions correspond to Bianchi Identities. We can then define
\be
\hat \Delta_{\bar{\mathbb{A}} \bar{\mathbb{B}}
\bar{\mathbb{C}}}{}^{\bar{\mathbb{D}}}= \left([\hat {\cal
L}_{\mathbb{E}_{\bar{\mathbb{A}}}}, \hat {\cal
L}_{\mathbb{E}_{\bar{\mathbb{B}}}}] \mathbb{E}_{\bar{\mathbb{C}}} - \hat {\cal
L}_{\hat {\cal L}_{\mathbb{E}_{\bar{\mathbb{A}}}}
\mathbb{E}_{\bar{\mathbb{B}}}}\mathbb{E}_{\bar{\mathbb{C}}}
\right){}^\mathbb{D} (\mathbb{E}^{-1})_{{\mathbb{D}}}{}^{\bar {\mathbb{D}}}= 0
\label{BIs}
\ee
More explicitly this is
\be
 \hat \Delta_{\bar{\mathbb{D}}\bar{\mathbb{A}}\bar{\mathbb{B}}}{}^{\bar
{\mathbb{C}}} =
 \hat \Delta_{\bar{E_\mathbb{D}}} \mathbb{F}_{\bar{
\mathbb{A}}\bar{\mathbb{B}}}{}^{\bar{\mathbb{C}}} =
 \left[\mathbb{F}_{\bar{\mathbb{D}}},
\mathbb{F}_{\bar{\mathbb{A}}}\right]_{\bar{\mathbb{B}}}{}^{\bar{\mathbb{C}}} +
 \mathbb{F}_{\bar{{\mathbb{D}}\bar{\mathbb{A}}}}{}^{\bar{\mathbb{E}}}
\mathbb{F}_{\bar {\mathbb{E}}\bar{\mathbb{B}}}{}^{\bar{\mathbb{C}}} - 2
\partial_{[\bar{\mathbb{D}}}
\mathbb{F}_{\bar{\mathbb{A}}]\bar{\mathbb{B}}}{}^{\bar{\mathbb{C}}}
 - \partial_{\bar{\mathbb{B}}} \mathbb{F}_{\bar
{\mathbb{D}}\bar{\mathbb{A}}}{}^{\bar{\mathbb{C}}}
 +
W^{\bar{\mathbb{C}}}{}_{\bar{\mathbb{B}}}{}^{\bar{\mathbb{E}}}{}_{\bar{\mathbb{F
}}} \partial_{\bar{\mathbb{E}}}
\mathbb{F}_{\bar{\mathbb{D}}\bar{\mathbb{A}}}{}^{\bar{\mathbb{F}}} = 0\nn
 \ee
These generalized BI include those of the four dimensional Riemann tensor and
that of the curvature for the two form (for simplicity we compute only the
projection of the latter)
\bea
\hat \Delta_{\bar d\bar a\bar b}{}^{\bar c} &=& 3 e_{\bar a}{}^\mu e_{\bar
d}{}^\nu e_{\bar b}{}^\rho (e^{-1})_\sigma{}^{\bar c} R_{[\mu\nu\rho]}{}^\sigma
= 3 (\partial_{[\bar a} \omega_{\bar d\bar b]}{}^{\bar c} + \omega_{[\bar a\bar
d}{}^{\bar e} \omega_{\bar b]\bar e}{}^{\bar c})
\\
\hat \Delta_{\bar d\bar a\bar b}{}^{\bar C} &=& e_{\bar d}{}^\mu e_{\bar
a}{}^\nu e_{\bar b}{}^\rho (\Phi^{-1})_M{}^{\bar C}\ (3 D_{[\mu} {\cal
F}_{\nu\rho]}{}^M - {\cal H}_{\mu\nu\rho}{}^M)
\eea
but more generally encode all other possible BI. For example, pursuing with
these computations one should obtain all the BIs of \cite{Greitz:2013pua}.

\section{Generalized diffeomorphisms in extended geometry}
\label{sec:extendedst}

In this section we address the construction of a full generalized Lie
derivative in the extended mega-space-time associated to the U-duality groups in
M-theory.  Using the particular $E_{7(7)}$ case analyzed in the previous section
as a guide line, here we will start general and do not specify any particular
duality group. Beginning with the canonical extension of a Lie derivative under
a Kaluza-Klein decomposition to the U-duality case, we compute the closure and
its failure. This allows to explore the extensions that form a closed algebra,
the appearance of the tensor hierarchy and the role of intertwiners.

\subsection{The canonical extension and the failure of closure}

We could start with a Lie derivative in higher-dimensions ($60$-in the $E_{7(7)}$ case) and split indices in
``external'' and ``internal'' components, namely $(\hat
V^\mu(x,Y), \hat V^M(x,Y))$. Now the vectors depend on ``external'' $x$ and
``internal'' $Y$ coordinates, and to make a difference with the vectors of
the previous section, here we put a hat on them.  Although we name the indices
as internal and external, let us emphasize that this just corresponds to an
index splitting, and here we are not assuming any compactification-type ansatz.
In
components it takes the form
\bea
(\hat {\cal L}_{\hat \xi} \hat V)^\mu &=& (L_{\hat \xi} \hat V)^\mu + ({\cal
L}_{\hat \xi} \hat V)^\mu -
({\cal L}_{\hat V} \hat \xi)^\mu\nn\\
(\hat {\cal L}_{\hat \xi} \hat V)^M &=& L_{\hat \xi} \hat V^M -
\hat V^\rho \partial_\rho \hat \xi^M + ({\cal L}_{\hat \xi} \hat V)^M
\label{GenLieDer1}
\eea
While in a conventional Kaluza-Klein splitting ${\cal L}_{\hat \xi} \hat V$
would correspond to the usual internal Lie derivative, here we promote it to
the internal extended generalized Lie derivative
\bea
({\cal L}_{\hat \xi} \hat V)^\mu &=& \hat \xi^P \partial_P \hat V^\mu +
\lambda \partial_P \hat \xi^P \hat V^\mu \nn\\
({\cal L}_{\hat \xi} \hat V)^M  &=& \hat \xi^P \partial_P \hat V^M - \hat V^P
\partial_P \hat \xi^M + Y^M{}_N{}^P{}_Q \partial_P \hat \xi^Q \hat V^N
\eea
This internal part was introduced in \cite{WaldramE,Berman:2012vc}, and here we
are allowing the external part to carry a weight $\lambda$. By now we will
remain general and do not select a particular $Y$-tensor, later we will
specialize to different cases.

We know that when this extended generalized Lie derivative is restricted purely
to the internal sector, the closure is achieved up to the so-called closure
constraints \cite{nosotros}. In particular, under a SS compactification, these
reproduce the quadratic constraints for the gaugings in gauged maximal
supergravity. On the other hand, the purely external sector closes
automatically, since it is governed by the $4$-dimensional Lie derivative of
Riemannian geometry. The combined case is however more tricky, and we can
already state that it will not close. In fact, under a generalized
Scherk-Schwarz dimensional reduction it reduces to (\ref{firstgenlie}), which
as we saw required an extension.

Let us however proceed with the computation of closure, to see what the failure
looks like in this more general case.  After some algebra, we find the
following result for the external components
\bea
(\hat \Delta_{\hat \xi_1} \hat {\cal L}_{\hat \xi_2} \hat V)^\mu &=&
Y^P{}_M{}^Q{}_N \left[\partial_Q\hat \xi_1^N \hat \xi_2^M \partial_P \hat V^\mu
+  \lambda  \partial_P(\partial_Q \hat \xi_1^N \hat \xi_2^M) \hat
V^\mu\right.\nn\\
&& \left. \ \ \ \ \ \ \ \ \ \ \ \ \ \ +2 \left(\partial_P \hat
\xi^\mu_{[1}\partial_Q \hat \xi^N_{2]}\hat V^M + \lambda  \partial_P
(\partial_Q \hat \xi^N_{[2} \hat V^M) \hat \xi_{1]}^\mu\right)\right]\nn\\
&& +\  \lambda  \left[(\partial_P\hat \xi_1^P) (L_{\hat \xi_2} \hat V)^\mu +
(\partial_P \hat V^P)(L_{\hat \xi_1} \hat \xi_2)^\mu + (\partial_P \hat
\xi_2^P) (L_{\hat V} \hat \xi_1)^\mu\right.\nn\\
&& \ \ \ \ \ \ \ +\ \hat V^\mu(\partial_P \hat \xi_1^\rho \partial_\rho \hat
\xi_2^P - \partial_P \hat \xi_2^\rho \partial_\rho \hat \xi_1^P)  + \hat
\xi_2^\mu (\partial_P \hat V^\rho \partial_\rho \hat \xi_1^P  - \partial_P \hat
\xi_1^\rho \partial_\rho \hat V^P) \nn\\
&& \ \ \ \ \ \ \ \left. +\ \hat \xi_1^\mu (\partial_P \hat \xi_2^\rho
\partial_\rho \hat V^P - \partial_P \hat V^\rho \partial_\rho \hat \xi_2^P)
\right]\label{closureexternal}
\eea
and the corresponding result for the internal components
\bea
(\hat \Delta_{\hat \xi_1} \hat {\cal L}_{\hat \xi_2} \hat V)^M &=& 2
Y^M{}_N{}^P{}_Q \hat V^Q\partial_P \hat \xi_{[1}^\rho \partial_\rho \hat
\xi_{2]}^N  - \hat V^\rho \left[ {\cal L}_{\partial_\rho \hat \xi_1} \hat \xi_2
+ {\cal L}_{\hat \xi_2} \partial_\rho\hat \xi_1\right]{}^M\nn\\
&& - \left[\left([{\cal L}_{\hat \xi_1} , {\cal L}_{\hat \xi_2} ] - {\cal
L}_{{\cal L}_{\hat \xi_1}\hat \xi_2}\right)\hat V\right]{}^M\nn\\
&& + \  \lambda  \left[ (\partial_P \hat \xi_1^P) (\hat \xi_2^\rho
\partial_\rho \hat V^M - \hat V^\rho \partial_\rho \hat \xi_2^M)  + (\partial_P
\hat \xi_2^P) (\hat V^\rho \partial_\rho \hat \xi_1^M - \hat
\xi_1^\rho\partial_\rho \hat V^M) \right.\nn\\
&& \left. \ \ \ \ \ \ \ +\ (\partial_P \hat V^P) (\hat \xi_1^\rho \partial_\rho
\hat \xi_2^M - \hat \xi_2^\rho\partial_\rho \hat \xi_1^M)
\right]\label{closureinternal}
\eea

Let us now analyze these results. We have split the external components
(\ref{closureexternal}) in two blocks. The first one vanishes by imposing the
internal closure constraints, but the second one does not. However, the latter
is proportional to $\lambda$, and then we can guarantee closure for the
external components provided we take
\be
\lambda = 0\label{lambda0}
\ee
We will assume this from now on. Then, in the internal components
(\ref{closureinternal}) the last block vanishes. The second line in
(\ref{closureinternal}) corresponds to the closure of the internal sector. It
was shown in  \cite{nosotros}  that this vanishes either by imposing the
section condition (\ref{sec1}), or the internal closure constraints implemented
in
Scherk-Schwarz reductions (see for example (\ref{falla1})). The first line is
however problematic, it does not vanish due to the section condition, nor (as we
saw) upon imposing the quadratic constraints in Scherk-Schwarz reductions. In
fact, under such a compactification ansatz while the first term compactifies to
zero
(we assume that under a compactification the external components depend on
external coordinates only), the second term compactifies to an
intertwining-dependent term, like that of (\ref{falla1}).

Then, for weight zero external components (\ref{lambda0}) and restricted
vectors (either due to the section condition or due to the more relaxed internal
closure constraints) the generalized Lie derivative (\ref{GenLieDer1}) fails to
close up to the following terms
\bea
(\hat \Delta_{\hat \xi_1} \hat {\cal L}_{\hat \xi_2} \hat V)^\mu &=&
0\label{failureclosure}\\\nn
(\hat \Delta_{\hat \xi_1} \hat {\cal L}_{\hat \xi_2} \hat V)^M
&=& 2\ Y^M{}_P{}^N{}_Q \partial_N\hat \xi_{[1}^\mu \partial_\mu
\hat \xi_{2]}^Q \hat V^P- \hat V^\mu Y^M{}_{P}{}^N{}_{Q} \partial_N
(\partial_\mu \hat \xi_1^Q \hat \xi_2^P)\\\nn
&& +\ 2\ \hat V^\mu Y^M{}_{[P}{}^N{}_{Q]}
\partial_\mu \hat \xi_1^Q \partial_N \hat \xi_2^P \nn
\eea
and then implies, as expected, that the extended generalized Lie derivative
must be completed.

\subsection{Including the tensor hierarchy}

 The structure of the failure  (\ref{failureclosure})  suggests the way in
which the generalized Lie derivative must be completed. In particular, the structure of the first
two terms suggest introducing the underlined terms
\bea
(\hat {\cal L}_{\hat \xi_1}\hat \xi_2)^\mu &=& (L_{\hat \xi_1}\hat \xi_2)^\mu +
\hat \xi_1^P
\partial_P  \hat \xi_2^\mu -  \hat \xi_2^P \partial_P  \hat \xi_1^\mu\nn\\
( \hat {\cal L}_{ \hat \xi_1} \hat \xi_2)^M &=&  L_{\hat \xi_1}\hat \xi_2^M -
\hat \xi_2^\rho \partial_\rho  \hat \xi_1^M + ({\cal L}_{\hat \xi_1}\hat
\xi_2)^M  \nn\\
&& + \frac12 \, \underline{Y^M{}_P{}^N{}_Q \partial_N  \hat \xi_{1\rho}{}^{PQ}
\hat \xi_2^\rho} + \frac12 \,\underline{Y^M{}_P{}^N{}_Q  \hat
\xi_{2\rho}{}^{PQ} \partial_N \hat \xi_1^\rho}\label{CompletedGenLie}
\eea
We have included new parameters $\hat \xi_\rho{}^{PQ}$, but note that in both
terms they are projected by the $Y$-tensor. This tensor selects the relevant
representations in the different duality groups. Let us anticipate the result:
this extension works for all $E_{n+1(n+1)}, n<6$ but
it fails for $E_{7(7)}$.

If one computes closure with this generalized Lie derivative,
without assuming any particular form for the invariant tensor $Y$, the result
for the internal sector reads
\bea
( \hat \Delta_{ \hat \xi_1}  \hat {\cal L}_{ \hat \xi_2}  \hat V)^\mu &=&
\frac 1 2 Y^P{}_R{}^T{}_S \left[\left(2 \partial_T \hat \xi_1^S \hat \xi_2^R +
\hat
\xi_2^\rho \partial_T \hat \xi_{1\rho}{}^{RS} + \hat \xi_{2\rho}{}^{RS}
\partial_T\hat \xi_1^\rho\right) \partial_P \hat V^\mu \right.  \\
&& \ \ \ \ \ \ \ \ - \left(2\partial_T \hat \xi_1^S \hat V^R + \hat V^\rho
\partial_T \hat \xi_{1\rho}{}^{RS} + \hat V_{\rho}{}^{RS} \partial_T\hat
\xi_1^\rho\right) \partial_P \hat \xi_2^\mu \nn\\
&&  \ \ \ \ \ \ \ \  - \left.\left(2\partial_T \hat \xi_2^S \hat V^R + \hat
V^\rho \partial_T \hat \xi_{2\rho}{}^{RS} + \hat V_{\rho}{}^{RS} \partial_T\hat
\xi_2^\rho\right) \partial_P \hat \xi_1^\mu \right]\nn
\eea
Every single term here is of the form $Y^P{}_R{}^T{}_S \ \partial_P \otimes
\partial_T$, so this vanishes under the section condition (\ref{sec1}), but
more generally
these correspond to closure constraints that compactify to zero.

Moving to the internal sector, after a long computation the closure of the
algebra can be taken to the form
\bea
( \hat \Delta_{ \hat \xi_1}  \hat {\cal L}_{ \hat \xi_2}  \hat V)^M &=&  \frac
1 2 \hat
V^\mu Y^M{}_P{}^N{}_Q \partial_N \left[( \hat \delta_{ \hat \xi_1} \hat
\xi_2)_\mu{}^{PQ} - L_{\hat \xi_1} \hat \xi_{2\mu}{}^{PQ} + 2 \hat \xi_2^\rho
\partial_{[\rho} \hat  \xi_{1\mu]}{}^{PQ} -2 \partial_\mu  \hat \xi_1^Q  \hat
\xi_2^P\right. \nn \\
&& \left. \ \ \ \ \ \ \ \ + 2 Y^P{}_R{}^T{}_S  \hat \xi_{[2}^Q \partial_T  \hat
\xi_{1]\mu}{}^{RS} + \partial_T \left(Y^P{}_R{}^T{}_S \hat \xi_1^Q \hat
\xi_{2\mu}{}^{RS} - \hat \xi_1^T \hat \xi_{2\mu}{}^{PQ}\right)\right]\nn\\
&& - \partial_N  \hat \xi_{[1}^\mu Y^M{}_P{}^N{}_Q \left[( \hat \delta_{ \hat
\xi_2]} \hat V)_\mu{}^{PQ} - L_{\hat \xi_{2]}}  \hat V_{\mu}{}^{PQ}  +  2 \hat
V^\rho \partial_{[\rho}  \hat \xi_{2]\mu]}{}^{PQ}- 2\partial_\mu  \hat
\xi_{2]}^Q  \hat V^P\right.\nn\\
&& \left. +  Y^P{}_R{}^T{}_S ( \hat V^Q \partial_T  \hat \xi_{2]\mu}{}^{RS}-
\hat \xi_{2]}^Q \partial_T  \hat V_\mu{}^{RS} )+ \partial_T
\left(Y^P{}_R{}^T{}_S \hat \xi_{2]}^Q \hat V_\mu{}^{RS} - \hat \xi_{2]}^T \hat
V_\mu{}^{PQ}\right)\right] \nn\\
&&+  Y^M{}_{[P}{}^N{}_{Q]} \left[ \hat  V^\mu Y^P{}_R{}^T{}_S
\left(2\partial_N\partial_T \hat \xi_{[2\mu}{}^{RS} \hat  \xi_{1]}^Q -
\partial_N\partial_T ( \hat \xi_1^Q  \hat \xi_{2\mu}{}^{RS})\right)\right. \nn\\
&&\left. +2 Y^P{}_R{}^T{}_S \left( \hat V^Q\partial_T  \hat \xi_{[2\rho}{}^{RS}
\partial_N \hat \xi_{1]}^\rho + 2  \hat V_\rho{}^{RS} \partial_{(N}  \hat
\xi_{[2}^Q \partial_{T)} \hat \xi_{1]}^\rho \right)+ 2\hat V^\mu \partial_\mu
\hat \xi_1^Q \partial_N \hat \xi_2^P \right]\nn\\
&& + \hat \Delta_{(SC)}^M \label{ClosureGeneral}
\eea
where we have collected all the terms that would vanish under the section
condition in the last term
\bea
\hat \Delta_{(SC)}^M &=& \frac 1 2 \left[Y^M{}_Q{}{}^{(N|}{}_P
Y^P{}_R{}^{|T)}{}_S -
Y^M{}_R{}^{(N}{}_S \delta^{T)}_Q \right]\label{DeltaSC}\\
&& \left(2 \hat V^Q \partial_T  \hat \xi_{2\rho}{}^{RS} \partial_N  \hat
\xi_1^\rho +  \hat V^Q \partial_T\partial_N  \hat \xi_1^\rho  \hat
\xi_{2\rho}{}^{RS} + 4 \partial_N  \hat \xi_{[2}^Q \partial_T \hat
\xi_{1]}^\rho  \hat V_\rho{}^{RS}\right. \nn\\
&& \left. +  \hat \xi_2^\rho \hat  V^Q \partial_T\partial_N \hat
\xi_{1\rho}{}^{RS} -  \hat V^\rho  \hat \xi_{[2}^Q \partial_T\partial_N  \hat
\xi_{1]\rho}{}^{RS} -  \hat V^\rho \partial_T\partial_N ( \hat \xi_1^Q  \hat
\xi_{2\rho}{}^{RS})\right)\nn\\
&& + \frac 1 2 Y^P{}_R{}^T{}_S \left(( \hat \xi_2^\rho \partial_T  \hat
\xi_{1\rho}{}^{RS} +  \hat \xi_{2\rho}{}^{RS} \partial_T  \hat
\xi_1^\rho)\partial_P  \hat V^M + 2 \hat V_{\rho}{}^{RS}\partial_T  \hat
\xi_{[2}^\rho \partial_P  \hat \xi_{1]}^M\right)\nn\\
&& +\left(\hat \Delta_{\hat \xi_1}\hat {\cal L}_{\hat \xi_2} \hat
V\right){}^M_{(i)}
\eea
The last term denoted by an $(i)$ contains the closure conditions of the
internal sector,
discussed in \cite{nosotros}.

We will now discuss this in detail for specific duality groups.

\subsubsection{The T-duality case: Double Field Theory}
When the duality group is  $O(n,n)$, namely T-duality, the $Y$-tensor is given
by
\be
Y^M{}_P{}^N{}_Q = \eta^{MN}\eta_{PQ}
\ee
where $\eta^{MN}$ is the symmetric duality invariant metric. In this case, the
generalized Lie derivative (\ref{CompletedGenLie}) reduces to
\bea
(\hat {\cal L}_{\hat \xi_1}\hat \xi_2)^\mu &=& (L_{\hat \xi_1}\hat \xi_2)^\mu +
\hat \xi_1^P\partial_P  \hat \xi_2^\mu -  \hat \xi_2^P \partial_P  \hat
\xi_1^\mu\\
( \hat {\cal L}_{ \hat \xi_1} \hat \xi_2)^M &=&  L_{\hat \xi_1}\hat \xi_2^M -
\hat \xi_2^\rho \partial_\rho  \hat \xi_1^M + ({\cal L}_{\hat \xi_1}\hat
\xi_2)^M  + \hat \xi_2^\rho\partial^M  \hat \xi_{1\rho} +
\hat \xi_{2\rho} \partial^M \hat \xi_1^\rho\nn
\eea
Here we have defined\footnote{This parameter should not be confused with the
4-dimensional vector $\hat \xi^{\mu}$, they correspond to completely different
gauge
parameters (the latter corresponds to $4$-dimensional diffeomorphisms, while
(\ref{xi_mu})
to gauge transformations of $B_{\mu\nu}$, as we will see shortly).}
\be \label{xi_mu}
\hat \xi_\mu = \frac 1 2\eta_{PQ}\ \hat \xi_\mu{}^{PQ}
\ee
since the $Y$-tensor selects its pure trace, and then leaves a unique 1-form
gauge parameter, and
\be
({\cal L}_{\hat \xi_1}\hat \xi_2)^M=  \hat \xi_1^P
\partial_P \hat \xi_2^M - \hat \xi_2^P \partial_P \hat \xi_1^M +  \partial^M
\hat \xi_{1P} \hat \xi_2^P
\ee
which is now the usual internal generalized Lie derivative of DFT. In
(\ref{ClosureGeneral}) all the last terms cancel for this
$Y$-tensor, and the only terms left (up to contributions that would vanish due
to the section condition) are the first four lines
 \bea
( \hat \Delta_{ \hat \xi_1}  \hat {\cal L}_{ \hat \xi_2}  \hat V)^\mu &=&  0 \\
( \hat \Delta_{ \hat \xi_1}  \hat {\cal L}_{ \hat \xi_2}  \hat V)^M &=&   \hat
V^\mu \partial^M \left[( \hat \delta_{ \hat \xi_1} \hat \xi_2)_\mu - (L_{\hat
\xi_1}\hat \xi_2)_{\mu} +  2\hat \xi_2^\rho\partial_{[\rho}  \hat
\xi_{1\mu]}\right.\nn\\
&&\ \ \ \ \ \ \ \ \ \ \  \left. - \partial_\mu  \hat \xi_{1P}  \hat \xi_2^P +
 \hat \xi_{2}^P \partial_P  \hat \xi_{1\mu}-\hat \xi_{1}^P \partial_P  \hat
\xi_{2\mu}\right]\nn\\
&& - 2\partial^M  \hat \xi_{2}^\mu  \left[( \hat \delta_{ \hat \xi_1} \hat
V)_\mu -(L_{\hat
\xi_1}\hat V)_{\mu} +  2\hat V^\rho\partial_{[\rho}  \hat
\xi_{1\mu]}\right.\nn\\
&&\ \ \ \ \ \ \ \ \ \ \ \ \ \ \   \left.- \partial_\mu  \hat \xi_{1P}  \hat
V^P +  \hat V^P \partial_P  \hat \xi_{1\mu}- \hat \xi_{1}^P \partial_P  \hat
V_\mu \right]+(1\leftrightarrow 2) \nn
\eea
Setting this to zero selects the proper transformation rule for the $1$-form
leading to the generalized Lie derivative in DFT
\bea\label{GenLieDFT}
(\hat {\cal L}_{\hat \xi_1}\hat \xi_2)^\mu &=& (L_{\hat \xi_1}\hat \xi_2)^\mu +
\hat \xi_1^P\partial_P  \hat \xi_2^\mu -  \hat \xi_2^P \partial_P  \hat
\xi_1^\mu\\\nn
( \hat {\cal L}_{ \hat \xi_1} \hat \xi_2)^M &=&  L_{\hat \xi_1}\hat \xi_2^M -
\hat \xi_2^\rho \partial_\rho  \hat \xi_1^M + ({\cal L}_{\hat \xi_1}\hat
\xi_2)^M  + \hat \xi_2^\rho\partial^M  \hat \xi_{1\rho} +
\hat \xi_{2\rho} \partial^M \hat \xi_1^\rho\\\nn
(\hat {\cal L}_{ \hat \xi_1} \hat \xi_2)_\mu &=&  (L_{\hat
\xi_1}\hat \xi_2)_{\mu} -  2\hat \xi_2^\rho\partial_{[\rho}  \hat \xi_{1\mu]} +
\partial_\mu  \hat \xi_{1P}  \hat \xi_2^P -
 \hat \xi_{2}^P \partial_P  \hat \xi_{1\mu}+\hat \xi_{1}^P \partial_P  \hat
\xi_{2\mu}\nn
\eea
This is the full generalized Lie derivative because in this case the last
component of the closure conditions vanishes up to closure constraints
\be
( \hat \Delta_{ \hat \xi_1}  \hat {\cal L}_{ \hat \xi_2}  \hat V)_\mu =  0
\ee
Then, this is the end of the story here, only one additional one-form has to be
included in order to achieve closure. This extra $1$-form gauge parameter
implies that there is only one $2$-form field in DFT, which is none but the
Kalb-Ramond field $B_{\mu\nu}$. Interestingly,  since in this case the number
of $1$-forms
and vectors coincides, all the gauge parameters can be grouped into the bigger
duality group $O(d+n,d+n)$ of DFT, and the generalized Lie derivative can be
condensed as
\be
\left(\hat {\cal L}_{\hat \xi_1}\hat \xi_2\right){}^{\mathbb{M}} = \hat
\xi_1^{\mathbb{P}} \partial_{\mathbb{P}} \hat \xi_2^{\mathbb{M}} - \hat
\xi_2^{\mathbb{P}} \partial_{\mathbb{P}} \hat \xi_1^{\mathbb{M}} +
Y^{\mathbb{M}}{}_{\mathbb{P}}{}^{\mathbb{N}}{}_{\mathbb{Q}}
\partial_{\mathbb{N}} \hat \xi_1^{\mathbb{Q}} \hat \xi_2^{\mathbb{P}}
\ee
where we have noted $\hat \xi^{\mathbb{M}} = (\xi^\mu, \xi^M,\xi_\mu)$ and
$\partial_{\mathbb{M}} = (\partial_\mu, \partial_M, 0)$, and
$Y^{\mathbb{M}}{}_{\mathbb{P}}{}^{\mathbb{N}}{}_{\mathbb{Q}}=
\eta^{\mathbb{MN}}\eta_{\mathbb{PQ}}$, with $\eta^{\mathbb{MN}}$ the invariant
metric of $O(d+n,d+n)$. In DFT, one can incorporate derivatives with respect to
the form directions $\partial^\mu$, and constrain or completely eliminate their
dependence by imposing
an external $O(d,d)$ section condition. We expect that this is also the case in
the general case of $U$-duality groups, although in this paper we are setting
them to zero explicitly.

\subsubsection{The $E_{7(7)}$ case}\label{n=6}

The above analysis fails to work for $E_{7(7)}$. The problem can already be
tracked back to equation (\ref{failureclosure}). The reason is that the last
term, while vanishing in the cases of $O(n,n)$ and $E_{n+1(n+1)}$, $n<6$,
does not vanish when the $Y$-tensor is that of $E_{7(7)}$ because it contains an
antisymmetric piece
\be
Y^M{}_P{}^N{}_Q = -12 P^{MN}{}_{(PQ)} + \underline{\frac 1 2 \omega^{MN}
\omega_{[PQ]}}
\ee
While the two extra terms in (\ref{CompletedGenLie}) cancel the failure of the
first two terms in (\ref{failureclosure}), there is no obvious covariant
completion to
the generalized Lie derivative that would cancel the last term in
(\ref{failureclosure}).

In close relation to this fact, although we have been able to close the algebra
in the compactified case, by adding the extra term in
(\ref{extraterm}), there is no obvious covariant uplift for this contribution,
such that
it compactifies to the form (\ref{projection1}), namely, the new gauge
parameter contracted with the intertwining tensor. To be more specific, let us
note that the intertwining tensor $F_{(AB)}{}^C$ can be cast as follows in terms
of the twist matrix
\be
F_{(AB)}{}^C = Y^M{}_P{}^N{}_Q \partial_N U_{(A}{}^Q U_{B)}{}^P (U^{-1})_M{}^C
 \ee
 Then, starting from the last term in the third line of
(\ref{componentsgeneralform}), it should uplift to
 \be
 \xi_\mu{}^{AB} F_{(AB)}{}^C U_C{}^M = \frac 1 2 Y^M{}_P{}^N{}_Q \partial_N
\hat \xi_\mu{}^{PQ} +  Y^M{}_{[P}{}^N{}_{Q]} \partial_N U_A{}^Q U_B{}^P
\xi_\mu{}^{AB}\label{upliftE7}
 \ee
 where we assume that the Scherk-Schwarz ansatz is of the form $\hat
\xi_\mu{}^{PQ} = U_A{}^P U_B{}^Q \xi_\mu{}^{AB}$. Clearly the uplift should not
be $U$-dependent, and then we see again that the failure to uplift the
intertwining term in the reduced generalized Lie derivative is proportional to
$Y^M{}_{[P}{}^N{}_{Q]}$. Then, the result is that without adding a new contribution to
(\ref{CompletedGenLie}) or supplementing it with some additional constraints,
the closure fails to work up to terms proportional to
the antisymmetric part of the $Y$-tensor (\ref{ClosureGeneral}).

Already in \cite{Cederwall:2013naa} there were indications that the $E_{7(7)}$
case is special in this respect: while the divergence of the $1$-form gauge
parameters  is covariant (the connection contributions vanish) for
$E_{n+1(n+1)}$ with $n < 6$, this is not the case of $E_{7(7)}$. More concretely, for $n < 6$ one has
\be
Y^M{}_P{}^N{}_Q \nabla_N \xi_\mu{}^{PQ} = Y^M{}_P{}^N{}_Q \partial_N \xi_\mu{}^{PQ}
\ee
and then this expression is connection free. Then, one
possibility is that the extra-contributions in (\ref{ClosureGeneral}) should be
actually defined in terms of a covariant derivative. Another possibility is
implementing a {\it field} section condition. As we will see in the next
section, when filling the fundamental $\bf 56$ representation with the degrees
of freedom of $11$-dimensional maximal supergravity, only $28$ components can
be associated to gauge parameters of the theory. The rest correspond to gauge
parameters that transform the fields that couple to dual branes. Then, perhaps
in this case the closure requires a duality covariant constraint of the form
$\omega_{PQ}\xi_1^P \xi_2^Q = 0$, such that its solutions select a
$28$-dimensional section of the parameter space, which would cancel the
problematic last term in (\ref{failureclosure}). Also, it could be possible
that even without a constraint of this form, one should supplement the algebra
with duality relations between gauge parameters. Analyzing these possibilities
lies beyond the scope of this paper, we leave this here as an open problem.

\subsubsection{ The $E_{n+1(n+1)}$ case for $n < 6$}\label{n<6}

We now move to the other exceptional groups. In the case of $E_{n+1(n+1)}$ with
$n < 6$, the internal $Y$-tensor is symmetric in the pairs of upper and lower
indices $Y^{M}{}_{[P}{}^N{}_{Q]} = 0$. Then, in (\ref{ClosureGeneral}) the
fifth and sixth lines vanish, and the first four determine the gauge
transformation of the new components\footnote{ We used the identity \bea
Y^M{}_P{}^N{}_Q \!\!\!\!\!\!&& \!\!\!\!\!\! \partial_T\left(Y^P{}_R{}^T{}_S
\hat \xi_1^Q  \hat \xi_{2\mu}{}^{RS} -  \hat \xi_1^T  \hat
\xi_{2\mu}{}^{PQ}\right) \partial_N  \hat \varphi^\mu\nn\\
&=& \left[\vphantom{\frac 1 2}Y^M{}_P{}^N{}_Q \xi_1^Q (Y^P{}_R{}^T{}_S
\partial_T  \hat \xi_{2\mu}{}^{RS}) - {\cal L}_{\xi_1} (Y^M{}_P{}^N{}_Q \hat
\xi_{2\mu}{}^{PQ})\right.\nn\\
&&  \ \ \ - Y^T{}_R{}^N{}_S  \hat \xi_{2\mu}{}^{RS} \partial_T  \hat \xi_1^M +
Y^N{}_P{}^T{}_Q \partial_T  \hat \xi_1^Q Y^M{}_R{}^P{}_S  \hat
\xi_{2\mu}{}^{RS} \nn\\
&&\ \ \ + 2 \partial_T \hat  \xi_1^Q  \hat \xi_{2\mu}{}^{RS}
\left(Y^M{}_Q{}^{(N|}{}_P Y^P{}_R{}^{|T)}{}_S - Y^M{}_R{}^{(N}{}_S
\delta^{T)}_Q\right)\nn\\
&&\ \ \ \left.- 4 \partial_T  \hat \xi_1^Q  \hat \xi_{2\mu}{}^{RS}
Y^M{}_{[Q}{}^{(N|}{}_{P]} Y^P{}_R{}^{|T)}{}_S \vphantom{\frac 1 2}\right]
\partial_N \hat \varphi^\mu\nn
\eea where the last three lines contain section condition-like terms that
contribute to (\ref{DeltaSC}) to form closure constraints that compactify to the quadratic constraints in gauge maximal supergravity, and terms antisymmetric in $Y$ that are
irrelevant in this subsection.}
\bea
Y^M{}_P{}^N{}_Q\partial_N \left(\hat {\cal L}_{\hat \xi_1} \hat
\xi_2\right){}_\mu{}^{PQ}
&=& Y^M{}_P{}^N{}_Q \partial_N\left[ L_{\hat \xi_1}\hat \xi_{2\mu}{}^{PQ} + 2
\hat
\xi_2^\rho \partial_{[\mu} \hat \xi_{1\rho]}{}^{PQ} + 2\partial_\mu \hat \xi_1^Q
\hat \xi_2^P \right.  \\
&&\ \ \ \ \ \ \left.- 2 Y^P{}_R{}^T{}_S \hat \xi_{[2}^Q \partial_T \hat
\xi_{1]\mu}{}^{RS} +({\cal L}_{ \hat \xi_1}  \hat \xi_{2\mu}{}^{PQ}-
Y^P{}_R{}^T{}_S  \hat \xi_1^Q \partial_T  \hat
\xi_{2\mu}{}^{RS})\right]\nn
\eea
Notice that this expression is projected by the ``intertwining''
operator\footnote{ We call it intertwining operator since it plays the same
projection role as the intertwining tensors (\ref{intertwining}) in the
compactified case.}
\be
\frac 1 2 Y^M{}_P{}^N{}_Q \ \partial_N \ (\dots)^{PQ} \ .
\ee
We can then proceed as in the previous section, and remove the projection up to
terms that vanish under it
\bea
 \left(\hat {\cal L}_{\hat \xi_1} \hat \xi_2\right){}_\mu{}^{<MN>}
&=&  L_{\hat \xi_1}\hat \xi_{2\mu}{}^{<MN>} + 2 \hat
\xi_2^\rho \partial_{[\mu} \hat \xi_{1\rho]}{}^{<MN>} + 2Y^M{}_P{}^N{}_Q
\partial_\mu \hat \xi_1^Q
\hat \xi_2^P + \hat \Gamma_\mu{}^{<MN>} \ \ \ \ \ \ \ \
\label{GaugeTransfmuPQ}\\
&& - 2 Y^M{}_P{}^N{}_Q \hat \xi_{[2}^Q \partial_T \hat
\xi_{1]\mu}{}^{<PT>} +{\cal L}_{ \hat \xi_1}  \hat \xi_{2\mu}{}^{<MN>}-
Y^M{}_P{}^N{}_Q  \hat \xi_1^Q \partial_T  \hat
\xi_{2\mu}{}^{<PT>}\ \ \ \ \nn
\eea
We are using the notation that
\be
\xi_\mu{}^{<MN>} = Y^M{}_P{}^N{}_Q \xi_\mu{}^{PQ}
\ee
and have then included $\hat \Gamma_\mu{}^{<MN>}$ which satisfies the relations
\be
 \partial_N \hat \Gamma_\mu{}^{<MN>} = 0 \ , \ \ \ \
\hat \Gamma_\mu{}^{<MN>} \partial_N \hat \varphi =
0\label{relationsGamma}
\ee
where $\hat \varphi$ represents any field or gauge parameter.
 As happens in the reduced case, $\hat
\Gamma_\mu{}^{<MN>}$ must be determined by demanding closure for this new
component
\bea
 \left(\hat \Delta_{\hat \xi_1} \hat{\cal L}_{\hat \xi_2} \hat
V\right){}_\mu{}^{<MN>} = 0
\eea
and this will require the new contributions to include the gauge parameters
components of the next step of the hierarchy, and so on.

Notice that the last two terms in (\ref{GaugeTransfmuPQ}) compactify to the
($E_{n+1(n+1)}, n<6$ analog of the)
first term in the fifth line of (\ref{componentsgeneralform}), which is
proportional to the intertwining tensor of the second level. Since the second
term in that line is also proportional to this intertwining tensor (contracted
with the new $2$-form component of the gauge parameters), we can use the last
two
terms in (\ref{GaugeTransfmuPQ}) to determine what the next intertwining
operator will be. After some algebra one can show that for a symmetric
$Y$-tensor, the following identity holds
\bea
Y^M{}_P{}^N{}_Q\,\,\!\!\!\!\!\!&&\!\!\!\!\!\!\left(Y^P{}_R{}^T{}_S  \hat
\xi_1^Q
\partial_T  \hat \xi_{2\mu}{}^{RS} - {\cal L}_{ \hat \xi_1}  \hat
\xi_{2\mu}{}^{PQ}\right)\nn\\
&=&
\partial_T (\xi_1^Q \xi_{2\mu}{}^{RS}) \tilde Y^{MNT}{}_{QRS}
- 3 \xi_{2\mu}{}^{RS} \partial_T \xi_1^Q \tilde
Y^{MNT}{}_{(QRS)}
\label{identitynewintertwiner}
\eea
where we have defined 
\bea
\tilde Y^{MNT}{}_{QRS} = Y^M{}_P{}^N{}_Q Y^P{}_R{}^T{}_S -
Y^M{}_R{}^N{}_S \delta^T_Q\label{tildeY}
\eea
The last term in (\ref{identitynewintertwiner}) vanishes in the cases
$E_{n+1(n+1)}$ for $n<5$ \cite{Berman:2012vc}, but not for $n = 5$. Then, at
least for $n<5$ and based on (\ref{CompletedGenLie}) one can conjecture that
the $\hat\Gamma_\mu{}^{<MN>}$ will contain  contributions of the form
\bea
 \hat \Gamma_{\mu}{}^{<MN>} = \frac 1 3
\left(\xi_2^\rho \partial_T \xi_{1\mu\rho}{}^{<MNT>} +
\xi_{2\mu\rho}{}^{<MNT>}\partial_T \xi_1^\rho\right) + \dots
\eea
where
\be
\xi_{\mu\nu}{}^{<MNT>} = \tilde Y^{MNT}{}_{QRS}\ \xi_{\mu\nu}{}^{QRS}
\ee
and that the intertwining operator of the following level will read
\be
\frac 1 3 \tilde Y^{MNT}{}_{QRS}\ \partial_T(\ldots)^{QRS}
\ee
Notice that in the $O(n,n)$ case, the $\tilde Y$-tensor vanishes, and this
explains why the tensor hierarchy ends at the $1$-form.

Let us note that the $Y$-tensor selects the correct representations for the
$1$-form vectors. For the different U-duality groups $E_{n+1(n+1)}$ it reads
\cite{Berman:2012vc}
\bea
&& n = 2: \ \ \ \ \ Y^{i\alpha}{}_{l\delta}{}^{j\beta}{}_{k\gamma} = 4
\delta^{ij}_{kl} \delta^{\alpha\beta}_{\gamma\delta}\nn\\
&& n = 3: \ \ \ \ \ Y^{M}{}_P{}^N{}_Q = \epsilon^{iMN} \epsilon_{iPQ} \nn\\
&& n = 4: \ \ \ \ \ Y^{M}{}_P{}^N{}_Q = \frac 1 2 (\gamma^i)^{MN}
(\gamma_i)_{PQ} \nn\\
&& n = 5: \ \ \ \ \ Y^{M}{}_P{}^N{}_Q = 10 d^{MNR} \bar d_{PQR} \label{Ytensors}
\eea
where for $n = 2$, the $SL(3)$ indices take values $i,j = 1,2,3$ and the
$SL(2)$ indices take values $\alpha,\beta = 1, 2$, for $n = 3$ the
$SL(5)$ indices take values $i = 1,\dots,5$ and $M=[ij]$, for $n = 4$ the
$\gamma$-matrices
correspond to the $16\times 16$
MW representation of $SO(5,5)$, so $i = 1,\dots,10$, and for $n = 5$ the
$d$-tensor is the symmetric invariant of $E_{6(6)}$. Notice that this tensor
projects the $1$-form components of the gauge parameters to the following
representations
\bea
Y^M{}_P{}^N{}_Q \xi_\mu{}^{PQ}= \left\{ \begin{matrix} n = 2: \ \ 4
\xi_{\mu}{}^{[ij][\alpha\beta]} \ \ \ \ \ \ \ \ \ \ \ \ \ \ \ \ \ \ \ \  \ \ \
\   \ \ \ \ \ \ {\bf (3,1)}\\
\noindent n = 3: \ \ \epsilon^{iMN} \xi_{\mu i}  \ \ \ \ \ \ \ \ \ \  \ \ \ \ \
\ \ \ \  \ \ \ \ \ \  \ \ \ \ \ \  \ \ \ \ \ \ {\bf 5}\\
n = 4: \ \ \frac 1 2(\gamma^i)^{MN} \xi_{\mu i}  \ \ \ \ \ \ \ \  \ \ \ \ \ \ \
\ \ \  \ \ \ \ \ \  \ \ \ \ \ \ {\bf 10}\\
n = 5: \ \ 10 d^{MNR} \xi_{\mu R} \ \ \ \ \ \ \ \ \ \  \ \ \ \ \ \  \ \ \ \ \ \
 \ \ \ \ \ \ \ \ \overline{{\bf 27}}\\
n = 6: \ \ -12 (t_\alpha)^{MN} \xi_\mu{}^\alpha + \frac 1 2 \omega^{MN} \xi_\mu
\ \ \ \ {\bf 133 + 1}
 \end{matrix}\right.
\eea
Accordingly, we expect the $\tilde Y$-tensor (\ref{tildeY}) to be related to
the representations of the $2$-form gauge parameters.

~

In the $E_{6(6)}$ case the obstruction is related to the fact that there is no clear uplift
for the ($E_{6(6)}$ analog of the) last component of the fifth line in
(\ref{componentsgeneralform}). This obstruction in the second level of
$E_{6(6)}$ has the same origin of the obstruction at the first level of
$E_{7(7)}$, discussed around equation (\ref{upliftE7}), and the way to circumvent them should proceed in the same way as in the first level of $E_{7(7)}$. Notice however that
since the first level of the algebra closes in this case, one can write an
action that includes the $2$-form curvature of the gauge fields, as in
\cite{HohmSamt}.

~

Let us now introduce the field degrees of freedom and compute the fluxes to see what the implications of this will be in the
construction of EFTs. Consider a generalized field-dependent frame, in the spirit of
(\ref{GenBein1})
\be
\hat {\mathbb{E}}_{\bar{\mathbb{A}}}{}^{\mathbb{M}} = \left(\begin{matrix}
e_{\bar{a}}{}^{\mu}
&-e_{\bar{a}}{}^{\rho}A_{\rho}{}^M &
 e_{\bar{a}}{}^{\rho}(B_{\rho\mu}{}^{<MN>} -
A_\rho{}^{<M}A_\mu{}^{N>})  \\
0 & \Phi_{\bar{A}}{}^M & 2 A_\mu{}^{<M} \Phi_{\bar{A}}{}^{N>}
 \\
0 & 0 & (e^{-1})_{\mu}{}^{\bar{a}}
\Phi_{\bar{A}}{}^{<M}\Phi_{\bar{B}}{}^{N>}    \\
   &   &   &  \ddots
\end{matrix}\right)\label{framegene}
\ee
and define the generalized fluxes as
\be
\hat {\mathbb{F}}_{ \bar{\mathbb{A}}  \bar{\mathbb{B}}}{}^{ \bar{\mathbb{C}}} =
(\hat
{\cal L}_{ \hat {\mathbb{E}}_{\bar{\mathbb{A}}}} \hat{\mathbb{
E}}_{\bar{\mathbb{B}}})^\mathbb{M} (\hat{\mathbb{E}}^{-1})_{{\mathbb{M}}}{}^{
\bar{\mathbb{C}}}
\ee
Just to put an example, they contain the $2$-form curvature for the gauge fields
\be
\hat{\mathbb{F}}_{\bar a \bar b}{}^{\bar C} = -e_{\bar a}{}^\mu e_{\bar
b}{}^\nu (\Phi^{-1})_M{}^{\bar C} {\cal F}_{\mu\nu}{}^M
\ee
with
\be
{\cal F}_{\mu\nu}{}^M = 2 \partial_{[\mu} A_{\nu]}{}^M - [[A_\mu,A_\nu]]^M -
\frac 1 2 Y^M{}_P{}^N{}_Q \partial_N B_{\mu\nu}{}^{PQ} \label{gaugecurvature}
\ee
where we have defined the internal exceptional C-bracket
\be
[[A_\mu , A_\nu]] = \frac 1 2 \left({\cal L}_{A_\mu} A_{\nu} - {\cal L}_{A_\nu}
A_{\mu}\right)
\ee
and assumed that the $Y$-tensor is symmetric. Since, at this level the algebra
closes for $E_{n+1(n+1)}, n<6$, it is possible to define a theory in terms of
this curvature, as achieved in \cite{HohmSamt} for $E_{6(6)}$.

The generalized frame above was written in a gauged fixed
$H=SO(1,d-1)\times
H_i$ triangular form ($H_i$ is the local compact
maximal subgroup
of the U-duality group). Then, in order to analyze the gauge transformations of its components, we must make sure that the generalized diffeomorphic transformations preserve this gauge choice, by performing a compensating $H$-transformation\footnote{ Let us emphasize that actually this is not needed in our
approach since by construction the frame
is indeed a collection of  vectors, and therefore its transformation is well
defined and independent of the choice of frame.}. The combined transformations that preserve the gauged-fixed form of the generalized frame read\footnote{ We note these transformation with $\delta'$, to distinguish them from the transformations $\delta$ considered before.}
\be
\delta' \hat {\mathbb{E}}_{\bar {\mathbb{A}}}{}^{\mathbb{M}} = \hat {\cal
L}_{\hat \xi} \hat {\mathbb{E}}_{\bar {\mathbb{A}}}{}^{\mathbb{M}} -
\Lambda_{\bar {\mathbb{A}}}{}^{\bar{\mathbb{B}}} \hat
{\mathbb{E}}_{\bar{\mathbb{B}}}{}^{\mathbb{M}}
\ee
We will focus for simplicity in the transformation of the vielbein $e_{\bar a}{}^\mu$, gauge fields $A_\mu{}^M$ and scalars $\Phi_{\bar A}{}^M$, but this analysis can be extended to the other components. In particular, demanding that the $()_{21}$ component remains zero under a gauge
transformation selects
\be
\Lambda_{\bar A}{}^{\bar b} = - (e^{-1})_\mu{}^{\bar b} \Phi_{\bar A}{}^P
\partial_P \hat\xi^\mu\ , \ \ \ \ \ \ \Lambda_{\bar a}{}^{\bar B} = \eta_{\bar
a \bar b} \delta^{\bar A\bar B} (e^{-1})_\mu{}^{\bar b} \Phi_{\bar A}{}^P
\partial_P \hat \xi^\mu
\ee
Notice that this compactifies to zero, which is the reason why no compensation of this form was required in the previous section.
 Here we introduced the
components of the invariant $SO(1,d-1)\times H_i$ metric, namely the flat
Minkowski and Euclidean metrics $\eta_{\bar a\bar b}$ and $\delta^{\bar A \bar
B}$. Then, we can readily find the gauge transformations of the components
\bea
\delta' e_{\bar a}{}^{\mu} &=& L_{\hat \xi} e_{\bar a}{}^\mu + \hat \xi^P
\partial_P e_{\bar a}{}^\mu + e_{\bar a}{}^\rho A_\rho{}^P \partial_P \hat
\xi^\mu \label{gaugetransfcomponentes}\\
\delta' A_\mu{}^M &=& L_{\hat \xi} A_\mu{}^M + \partial_\mu \hat \xi^M + {\cal
L}_{\hat \xi} A_\mu{}^M - \frac 1 2 \partial_N \hat \xi_\mu{}^{<MN>} \nn\\
&& - \frac 1 2 (B_{\mu\rho}{}^{<MN>} - A_\mu{}^{<M} A_{\rho}{}^{N>}) \partial_N
\hat \xi^\rho - A_\mu{}^P \partial_P \hat \xi^\rho A_\rho{}^M + g_{\mu\nu}
{\cal M}^{MP} \partial_P \hat \xi^\nu\nn\\
\delta' \Phi_{\bar A}{}^M &=& L_{\hat \xi} \Phi_{\bar A}{}^M + {\cal L}_{\hat
\xi} \Phi_{\bar A}{}^M + A_\rho{}^{<M} \Phi_{\bar A}{}^{N>} \partial_N \hat
\xi^\rho - A_\mu{}^M \Phi_{\bar A}{}^P \partial_P \hat \xi^\mu \nn
\eea
where we have defined the internal ``scalar'' matrix ${\cal M}^{MN} =
\Phi_{\bar A}{}^M \delta^{\bar A \bar B} \Phi_{\bar B}{}^N$ and the ``external''
metric $g_{\mu\nu} = (e^{-1})_\mu{}^{\bar a} \eta_{\bar a \bar b}
(e^{-1})_\nu{}^{\bar b}$.

For the sake of completeness, we now establish the redefinitions required to make contact with the results in \cite{HohmSamt}. Regarding the fluxes, one can already see that the gauge curvature (\ref{gaugecurvature}) coincides. Focussing on the internal transformation of the gauge field $A_\mu{}^M$, we find that redefining the $1$-form gauge parameter as
\be
\frac 1 2 \hat \xi_\mu{}^{<MN>}  =   \hat \Xi_\mu{}^{<MN>} +   \hat \xi^{<M} A_{\mu}{}^{N>}
\ee
one obtains
\be
\delta' A_{\mu}{}^M = {\cal D}_\mu \hat \xi^M - \partial_N \hat \Xi_\mu{}^{<MN>} \ , \ \ \ \ {\cal D}_\mu = \partial_\mu - {\cal L}_{A_\mu}
\ee
which is the internal gauge transformation of the gauge fields as exposed in \cite{HohmSamt}. There, instead of considering purely external diffeomorphisms, the authors combined them with a subsector of internal diffeos where the internal gauge parameters are restricted to the form $\hat \xi^M = - \hat \xi^\nu A_\nu{}^M$, such that this particular combination results in gauge transformations that look like external diffeomorphisms, but are manifestly covariantized with respect to the gauge fields. To make contact with those transformations, one also has to make the following redefinition
\be
\hat \Xi_\mu{}^{<MN>} \to \hat \Xi_\mu{}^{<MN>} - \frac 1 2 \hat \xi^\rho (B_{\mu\rho}{}^{<MN>} - A_\mu{}^{<M} A_\rho^{N>})
\ee
and then the gauge transformations (\ref{gaugetransfcomponentes}) reduce to
\bea
\delta' e_{\bar a}{}^\mu &=& \hat \xi^\rho {\cal D}_\rho e_{\bar a}{}^\mu - e_{\bar a}{}^\rho {\cal D}_\rho \hat \xi^\mu\nn\\
\delta' A_\mu{}^M &=& \hat \xi^\rho {\cal H}_{\rho\mu} + g_{\mu\nu} {\cal M}^{MP} \partial_P \hat \xi^\mu \\
\delta' \Phi_{\bar A}{}^M &=& \hat \xi^\rho {\cal D}_\rho \Phi_{\bar A}{}^M\nn
\eea
One could also extended this analysis for the $2$-form field, were we expect the transformation to depend on its curvature, and then it seems that duality relations would be required to make contact with \cite{HohmSamt}. It would be interesting to explore the role of dualities of that type at the level of the algebra of generalized diffeomorphisms discussed here.

\subsection{The gauge structure of the full generalized Lie derivative}

Collecting the information from the previous subsections, we can now try to
condense the notation. As in the compactified case, the gauge parameters can be
thought of as components of a generalized gauge parameter in an extended
mega-space
\be
\hat \xi^{\mathbb{M}} = (\hat \xi^\mu, \hat \xi^{M}, \hat \xi_\mu{}^{<MN>}, \hat
\xi_{\mu\nu}{}^{<MNP>},\dots)
\ee
One can then define a generalized Lie derivative that takes the form
\be
\left(\hat {\cal L}_{\hat \xi_1} \hat \xi_2\right){}^{\mathbb{M}} = \hat
\xi_1^{\mathbb{P}} \partial_{\mathbb{P}} \hat \xi_2^{\mathbb{M}} - \hat
\xi_2^{\mathbb{P}} \partial_{\mathbb{P}} \hat \xi_1^{\mathbb{M}} +
Y^{\mathbb{M}}{}_{\mathbb{P}}{}^{\mathbb{N}}{}_{\mathbb{Q}}
\partial_{\mathbb{N}} \hat \xi_1^{\mathbb{Q}}\hat \xi_2^{\mathbb{P}}
\label{fullgenlie}
 \ee
The derivatives are restricted to only have components $\partial_{\mathbb{M}} =
(\partial_\mu, \partial_M,0,\dots)$. One could include derivatives with respect
to all the coordinates, and impose a generalized section condition (this is
what happens in the DFT T-duality case, where the derivatives in the directions
of the $1$-form gauge parameters $\partial^\mu$ are included for full covariance
of the theory, and then removed through a section condition). The $Y$-tensor
contains $GL(d)$ and $E_{n+1(n+1)}$ invariants, and when it is restricted to
the internal sector it coincides with the Y-tensors defined in (\ref{Ytensor})
and (\ref{Ytensors}). The purely external components of it vanish, and then one
recovers standard Riemannian geometry in the purely external sector.

In components, this generalized Lie derivative takes the form
\bea
(\hat {\cal L}_{\hat \xi_1}\hat \xi_2)^\mu &=& L_{\hat \xi_1} \hat \xi_2^\mu +
\hat \xi_1^P \partial_P  \hat \xi_2^\mu -  \hat \xi_2^P \partial_P  \hat
\xi_1^\mu\label{fullcomponents}\\
( \hat {\cal L}_{ \hat \xi_1} \hat \xi_2)^M &=&  L_{\hat \xi_1}  \hat \xi_2^M -
 \hat \xi_2^\rho \partial_\rho  \hat \xi_1^M +  {\cal L}_{\hat \xi_1} \hat
\xi_2^M  + \frac 1 2  (\partial_N  \hat \xi_{1\rho}{}^{<MN>}  \hat
\xi_2^\rho +  \hat \xi_{2\rho}{}^{<MN>} \partial_N \hat \xi_1^\rho)\nn\\
 \left(\hat {\cal L}_{\hat \xi_1} \hat \xi_2\right){}_\mu{}^{<MN>} &=& L_{\hat
\xi_1}\hat \xi_{2\mu}{}^{<MN>} - 2 \hat \xi_2^\rho \partial_{[\rho} \hat
\xi_{1\mu]}{}^{<MN>} +  Y^M{}_P{}^N{}_Q (2 \partial_\mu \hat \xi_1^Q \hat
\xi_2^P -
  \hat \xi_{2}^Q \partial_T \hat \xi_{1\mu}{}^{<PT>})\nn\\
&& + {\cal L}_{ \hat \xi_1}  \hat \xi_{2\mu}{}^{<MN>}+ \frac 1 3
\left(\xi_2^\rho \partial_T \xi_{1\mu\rho}{}^{<MNT>} +
\xi_{2\mu\rho}{}^{<MNT>}\partial_T \xi_1^\rho\right)\nn\\
&\vdots&\nn
\eea

We have only worked the hierarchy up to this level, but the computations can be
pushed forward with more effort. The closure of the external  component
is achieved up to terms that would vanish under imposing the section
condition (and compactify to zero). The first level internal component includes terms
that are projected by the first intertwining operator
\be
Y^M{}_P{}^N{}_Q\ \partial_N (\dots)^{PQ} = \partial_N (\dots)^{<MN>}
\ee
and closes up to section-condition-like terms (that compactify to quadratic
constraints) and terms proportional to $Y^M{}_{[P}{}^N{}_{Q]}$. While this
vanishes when the U-duality group is $E_{n+1(n+1)}$ with $n<6$,  we find a
closure obstruction for $E_{7(7)}$, on which we commented in Subsection \ref{n=6}. Regarding the second level component $\hat \xi_\mu{}^{<MN>}$, it includes in its transformation terms that are projected by the
second intertwining operator
\be
\tilde Y^{MNT}{}_{QRS}\ \partial_T (\dots)^{QRT} = \partial_T
(\dots)^{<MNT>}
\ee
The closure obstruction in this case is proportional to $\tilde
Y^{MNT}{}_{(QRS)}$. Now this vanishes for $n<5$, but not for
$E_{6(6)}$, and this obstruction was commented in Subsection \ref{n<6}.  We expect issues like this will appear at level $7-n$ for
$E_{n+1(n+1)}$, for the adjoint representations, but this has to be explored further.

We have also shown how to introduce a field-dependent generalized frame and how to extract the gauge transformations on its components, and how to build the covariant quantities. These results will be useful when constructing full duality invariant Exceptional Field Theories.

\section{Contact with 11-dimensional supergravity (and beyond)}
\label{sec:11dsugra}

In this section we concentrate on $d=4$ and U-duality group $E_{7(7)}$. Using a
$GL(7)\subset SL(8) \subset
E_7$ decomposition, were $GL(7)$ acts on the ``internal"
tangent space in the supergravity limit, i.e. on the coordinates dual to momentum, which can be either defined by hand, or by a section condition. By this, we
give the 11-dimensional origin of the gauge parameters,
external $1$-forms and some of the $2$-forms, etc. that we have seen. The same
can be done for
type II theories, but we will skip it (all the information about the purely
internal objects can be found for example in \cite{Aldazabal:2010ef}).

The exceptional generalized tangent space encodes all the gauge transformations
of 11-dimensional supergravity, namely diffeomorphisms and gauge transformations of the
$3$-form field. In order to recover an electro-magnetic covariant formulation,
we need to include their duals, namely a $5$-form corresponding to gauge
transformations of the $6$-form potential, and a gauge transformation for a
``dual graviton", transforming as a $7$-form times a $1$-form \cite{PachecoWaldram}.
The generalized tangent space can equivalently be constructed by counting the
charges of the theory, namely momentum, M2 and M5 brane charge, and KK monopole
charge (being the electric charge for the dual graviton). Furthermore, to build
the tensor hierarchy, or equivalently to go higher levels, more brane charges
are needed, corresponding to the so-called ``U-branes" or ``exotic branes"
\cite{Hullexotic}-\cite{exoticbranes} (i.e. configurations
with non-trivial U-duality monodromies). For example, to reconstruct the $133$
space-time $1$-forms, we need to add the exotic branes termed $5^3$ and $2^6$ in
\cite{exoticbranes} (see \cite{Hullexotic} for an earlier discussion). These
have
no direct 11-dimensional interpretation, as they exist only in spaces with 3 and 6
isometries respectively, and are obtained from the type IIA NS5 brane by a
chain of dualities and uplift. They transform as an $8$-form tensor, a $3$-form and
$6$-form respectively, and appear already when counting scalar degrees of freedom
in $E_{8(8)}$ \cite{SC}.
The generalized tangent space is therefore locally
\beq
\begin{aligned}
E = \, &T& \oplus &\ \Lambda^2 T^* & \oplus &\ \Lambda^5 T^*& \oplus &\
\Lambda^7 T^*
\otimes T^* & \oplus &\ \Lambda^8 T^* \otimes  \Lambda^3 T^*& \oplus &\
\Lambda^8 T^* \otimes  \Lambda^6 T^*& \oplus  &\cdots &  \\
&p& &M2& &M5& &\ \ \ \ \ \ KK&  &\ \ \ \ \ \ \ \  5^3& &\ \ \ \ \ \ \ \  2^6 &
&\cdots &
\end{aligned}
\eeq
where $T$ is 11-dimensional. The dots account for higher p-form representations.
For example, to obtain the $912$ degrees of freedom that appear in space-time
$2$-forms, we need extra forms transforming as $9$-forms times some lower form.
These show up  in the $l_1$ representation of $E_{11}$ at fourth order
\cite{E11}.

Splitting $T=T_4 + T_7$ we get the $GL(4)$ and $GL(7)$ representations
indicated in Table\ref{ta:charges}.\footnote{There is a subtlety in counting
degrees of freedom: the ``dual branes" (KK, $5^3$, $2^6$, etc) exist only in
spaces with $U(1)$ isometries. Their respective  $T^*$, $\Lambda^3 T^*$ and
$\Lambda^6 T^*$ lie along these compact directions. We have therefore excluded
the possibility of placing 4-dimensional space-time indices along them. If on the contrary
we allow this possibility, the KK monopole would count as $50$ $1$-form degrees of
freedom, and the total number of $1$-forms would be $133+1$, in accordance with
the $E_{11}$ results \cite{E11}, while the total number of $2$-forms would be
$912+56$.}
\begin{table}[!ht]
\begin{center}
\begin{tabular}{|c|ccccccc|c|c|} \hline
 & $p$ & $M2$ & $M5$& $KK$ & $5^3$ &$2^6$ &$\cdots$& &  \\
 GL(4) & \multicolumn{7}{c|}{GL(7)} &{\rm total} & $E_{7(7)}$ repr \\ \hline \hline
\rm{scalar} & 7&21 &21&7&0 & 0 &0 &56 & $\xi^M \in {\bf 56}$  \\
\rm{1-form} & 0&7 &35 &49 &35&7 &0 & 133 &$ \xi_{\mu}{}^{\alpha} \in
{\bf 133}$ \\
\rm{2-form} & 0&1 &35&147&245 &49 &$\cdots$ & 477+$\cdots$ &$\xi_{\mu\nu}{}^{\cal
A} \in {\bf 912} $ \\
\rm{3-form} & 0&0 &21&245&735 & 147& $\cdots$  &
1148+$\cdots$&$\xi_{\mu\nu\rho}{}^{\bf A} \in {\bf 133 }+ {\bf 8645} $
\\ \hline
\end{tabular}
\caption{\small
$GL(4)$ and $GL(7)$ decomposition of the generalized tangent space. We give the
number of internal components.}\label{ta:charges}
\end{center}
\end{table}

The bosonic fields of 11-dimensional supergravity are the metric and the $3$-form gauge
field. Again, in order to achieve 4-dimensional electro-magnetic invariance, we need to
include their duals, a $6$-form field $A_6$ and the dual graviton $B$,
transforming as $\Lambda^8 T^* \otimes T^*$ \cite{SC}.
The ``gauge field" whose electric charge are
the exotic branes included so far, termed $A_{9^3}$ and $A_{9^6}$, should
transform respectively as $\Lambda^9 T^* \otimes \Lambda^3 T^*$ and $\Lambda^9
T^* \otimes \Lambda^6 T^*$.
Again, to fill
out the corresponding $E_{7(7)}$ representations at the level of space-time 3-forms
and on, we need to include higher p-forms.
The $GL(4)$ and $GL(7)$ representations we
obtain for the bosonic fields are   shown in table \ref{ta:fields}.
\begin{table}[!ht]
\begin{center}
\begin{tabular}{|c|ccccccc|c|c|} \hline
 &  $g$ & $A_3$ & $A_6$& $B$ & $A_{9^3}$ & $A_{9^6}$& $\cdots$ & &  \\
GL(4)  & \multicolumn{7}{c|}{GL(7)} & {\rm total} &$E_{7(7)}$ repr \\ \hline \hline
\rm{scalar} & 28&35 &7&0&0&0&$0$ & 70 &$\Phi_{\bar A}{}^{M} \in$ {\bf 133}  \\
\rm{1-form} & 7&21 &21&7&0 &0&$0$ & 56 &$A_{\mu}{}^{M} \in$ {\bf 56} \\
\rm{2-form} & 0&7 &35&49&35&7&0 & 133 &$B_{\mu\nu}{}^{\alpha} \in$  {\bf 133} \\
\rm{3-form} & 0&1 &35&147&245 & 49 &$\cdots$&477+$\cdots$
&$C_{\mu\nu\rho}{}^{\cal A} \in$  {\bf 912} \\
\rm{4-form} & 0&0 &21& 245&735 &147 &$\cdots$ & 1148+$\cdots$
&$D_{\mu\nu\lambda\rho}{}^{\bf A} \in$  {\bf 133 }+ {\bf 8645} \\ \hline
\end{tabular}
\caption{\small
$GL(4)$ and $GL(7)$ decomposition of the bosonic fields. $B$ represents the
dual graviton, and $A_{9^3}$, $A_{9^6}$ the gauge fields that couple
electrically to the $5^3$ and $2^6$ branes. }\label{ta:fields}
\end{center}
\end{table}

 It will be useful to find the  embedding of the metric and 3-form field, as
well as their gauge transformations (diffeomorphisms and 2-forms) in terms of
$SL(8)$ representations. For that, we
use the following notation: $a,b,...=1,..,8$ are indices in the fundamental of
$SL(8)$, while $m,n=1,...,7$ are in the fundamental of $GL(7)$. Later on we
will use $\tM=1,...,11$ to denote
indices along eleven-dimensional space-time (i.e. $\tM=(\mu,m)$).

The fundamental
 representation of $E_{7(7)}$ decomposes with respect to $SL(8)$ into
\begin{align}
\bf{56}&=\bf{28}+{\bf{28'}} \label{slf} \quad \\
\xi&=(\xi^{ab}, {\xi}_{ab})\nonumber \
\end{align}
and the adjoint
\begin{align}
\bf{133}&=\bf{63}+\bf{70}\label{sla} \\
\Phi&=(\Phi^{a}{}_{b},\Phi_{abcd}) \nonumber
\end{align}
where $\Phi^a{}_a=0$ and $\Phi_{abcd}$ is fully antisymmetric. Finally, the
${\bf 912}$ decomposes into
\begin{align}
\bf{912}&=\bf{36}+\bf{420}+\bf{36'}+\bf{420'}\label{sl912} \\
F &= (F^{ab},F^{abc}{}_{d}, F_{ab}, F_{abc}{}^{d}) \nonumber
\end{align}
where $F^{ba}=F^{ab}$ and $F^{abc}{}_c=0$ and similarly for the objects with
the indices down.

We can then embed the generators of diffeomorphisms $v$ and two-form gauge
transformations $\lambda_2$ of conventional 11-dimensional supergravity, whose different
external components are given in the first two columns in Table
\ref{ta:charges}, in the following way
\beq \label{gp11d}
\begin{aligned}
\xi^{m8}&=&v^m \ , \qquad \xi_{mn}&=&\lambda_{mn} \  & \in&{\bf 28}+{\bf 28}'
&\in&  {\bf 56}    \\
\xi^{\mu}&=&v^{\mu} \ , \qquad \xi_{\mu}{}^8{}_m&=&\lambda_{\mu m} \  & \in&
{\bf 1}+ {\bf 63} &\in& {\bf 1}+  {\bf 133} \\
&& \xi_{\mu\nu}^{88}&=&\lambda_{\mu\nu} \  & \in& {\bf 36} &\in& {\bf 912}
\end{aligned}
\eeq

The scalar, vector, $2$-form and $3$-form fields coming from the metric and $A_3$
are embedded as
\beq \label{gf11d}
\begin{aligned}
\Phi_m{}^n &=&a_m{}^n \ ,  \quad \Phi_{mnp8}&=&A_{mnp}   & \in&{\bf 70}&\in&
{\bf 133}    \\
A_{\mu}{}^{m8} &=& a_\mu{}^m \ , \quad  A_{\mu}{}{}_{mn}&=&A_{\mu mn}  & \in&
{\bf 28}+ {\bf 28}' &\in& {\bf 56} \\
&& B_{\mu\nu}{}^{8}{}_m &=&A_{\mu\nu m}   & \in& {\bf 63} &\in& {\bf 133} \\
&& C_{\mu\nu\rho}{}^{88} &=&A_{\mu\nu \rho}   & \in& {\bf 36} &\in& {\bf 912} \\
\end{aligned}
\eeq
Given this, the first row of the generalized vielbein (\ref{GenBein1}) for 11-dimensional
supergravity reads therefore
\bea
{\mathbb E}_{\bar a}{}^{\mathbb M}&=&- e_{\bar a}{}^\rho  \big(
-\delta_\rho^\mu, [a_{\rho}{}^m]^8+ A_{\rho mn},
[ A_{\rho \mu m} + a_{\rho}{}^p A_{p \mu m}]^8, [A_{\mu \nu \rho} +
a_{\rho}{}^p A_{\mu \nu p}+\tfrac13 a_{\rho}{}^p a_{\mu}^q A_{\nu pq}]^{88} ,
\nn \\
& & \qquad \quad [A_{\rho [mn} A_{p]\mu \nu} +\tfrac13 a_{\rho}{}^q A_{q \mu[m}
A_{np] \nu}+\tfrac13 A_{\rho mn} a_{[\mu}{}^q A_{\nu]pq}]^8 \big) \ ,
\eea
where the superindex 8 completes the $SL(8)$ representation (for example the
second term corresponds to ${\mathbb M}={}^{m8}$).

One can now use the generalized Lie derivative (\ref{CompletedGenLie}) with
parameters (\ref{gp11d}), applied to this generalized vielbein
and obtain the gauge transformations of the fields in 11-dimensional supergravity. After
some algebra, one can show that from $(\hat{\cal L}_{\hat \xi} \hat{\mathbb E}_{\bar
a})^{\mu}$ and $(\hat{\cal L}_{\hat \xi} \hat {\mathbb E}_{\bar a})^{M}$ one recovers
precisely the gauge transformations of 11-dimensional supergravity, namely
\bea
\delta e_{\bar a}{}^{\tM}&=&{\tL}_v e_{\bar a}{}^{\tM} \nn \\
\delta A_{\mu mn} &=& {\tL}_v A_{\mu mn} + ({\rm d} \lambda)_{\mu mn}
\eea
where  $\tL_v$ means the ordinary Lie derivative with parameter
$v^{\tM}=(v^\mu, v^m)$, and we have used $e_{\bar a}{}^m=e_{\bar a}{}^{\rho}
a_{\rho}{}^m$.

Note that we have correctly reproduced the gauge transformations of 11-dimensional
supergravity using the generalized Lie derivative (\ref{fullcomponents})
which, as we have shown, does not close at the first level for $E_{7(7)}$. This is so because the obstruction for
closure goes away when the gauge parameters and the fields are those of 11-dimensional
supergravity. Indeed, on the one hand they obviously satisfy the section condition.
On the other, terms involving $Y^M{}_{[P}{}^N{}_{Q]}=\tfrac12 \omega^{MN}
\omega_{PQ}$ go away when contracted with gauge fields or gauge parameters, as
the only non-zero components of the latter are those with fundamental indices
$P$ or $Q={}^{m8}$ or $P={}_{mn}$, and therefore vanish in the symplectic products.

Furthermore, it is not hard to show that
restricting the generalized tangent space to vectors and two-forms, i.e. for $U=(u, \lambda)$, $V=(v,\omega)$, $u,v \in T$, $\lambda, \omega \in \Lambda^2 T^*$, the generalized Lie derivative
\beq
{\cal L}_U V=(L_u v, L_u \omega - \iota_v {\rm d} \lambda)
\eeq
closes without any need to introduce other forms.

\section{Conclusions} \label{sec:Conclu}

We have addressed the construction of U-duality invariant generalized
diffeomorphisms for M-theory. Taking as a starting point the generalized Lie
derivatives for the different U-duality groups introduced in
\cite{WaldramE},\cite{Berman:2012vc}, we explored the completion to the full
external plus internal space-time.

We began with the Scherk-Schwarz-type compactification ansatz leading to
gauged maximal supergravity, and specialized the analysis to the $E_{7(7)}$ case
in $4$-dimensions in Section \ref{sec:E}. Closure of the algebra is highly
non-trivial and requires an extension of the tangent space to include the
so-called tensor hierarchy. The closure at each level of the hierarchy dictates
the transformation rules of the following level. There is also a hierarchy of
intertwining tensors, such that when a given level of the hierarchy is
projected by the corresponding intertwiner, the (projected) sub-algebra formed
by it and the previous levels closes.  Interestingly, this procedure allowed us
to build a full generalized Lie derivative in the reduced case, that encodes all
the gauge transformations of maximal supergravity and the tensor hierarchy
(\ref{generalform})-(\ref{componentsgeneralform}).

The field content is introduced through a generalized frame (\ref{GenBein1}),
in such a way that the different components transform appropriately
(\ref{componentsgaugetransf}). This is a novel construction, in which the
different levels of the tensor hierarchy are embedded into different components of a generalized frame,
in analogy to what happens with the $2$-form in DFT. We then define generalized
fluxes in terms of the generalized frame and Lie derivative (\ref{genfluxes}),
and show that the different components correspond to covariant quantities in
the theory (\ref{genfluxcomponents}): curvatures, covariant derivatives, etc.
Since the generalized Lie derivative closes, the closure of the generalized
fluxes define the BIs of the theory (\ref{BIs}).

Encouraged by this construction, we then moved towards a full generalized Lie
derivative on the  extended space (\ref{fullgenlie}). Again, closure requires
an extended tangent space to accommodate the tensor hierarchy, and level by
level the closure conditions dictate the transformation rules of the next level.
We identified the ``intertwining operators'' of the first levels, which play a
role analog to (and compactify to) the intertwining embedding tensor in the
reduced theory.

Intriguingly, we found closure obstructions at different levels of the
hierarchy for different U-duality groups. Although we have commented on how they
can be circumvented, this issue deserves further study.

Let us finally comment on future lines of investigation. Regarding the reduced
case of Section \ref{sec:E}, it would be interesting to construct a generalized
geometry through covariant derivatives, connections, torsion and curvatures.
This is expected to give rise to an action and equations of motion of a
democratic formulation of maximal gauged supergravity. This framework can also
be used to explore non-geometry, and possible uplifts for the new maximal
gauged supergravities in \cite{Dall'Agata:2012bb}. Regarding the extended
construction of Section \ref{sec:extendedst}, it would be interesting to further
explore mechanisms to achieve full closure, and apply the generalized frame and
flux techniques of section \ref{sec:E} to build covariant quantities and actions
of Exceptional Field Theories. A construction of an underlying
generalized geometry with generalized connections and curvatures  would allow to systematically obtain the actions and equations of motion.

\section*{Acknowledgments}
We thank  B. de Witt, J. Palmkvist, M. Shigemori, D. Waldram and specially A.
Coimbra for useful discussions and comments. G. A.
thanks the ICTP and CERN for hospitality during the completion of this work.
D. M. and J. A. R. thank Institut de Physique Th\'eorique,
CEA/Saclay, for hospitality.
This work was partially supported
by EPLANET, CONICET, PICT-2012-513 and the ERC Starting Independent Researcher
Grant 259133-ObservableString.

\end{document}